\documentclass[11pt,a4paper]{IEEEtran}

\usepackage{multicol}
\usepackage{amsmath}
\usepackage{amssymb}
\usepackage{graphicx}%
\usepackage{bm}%
\usepackage{color}%
\usepackage{multirow}
\usepackage{cite}
\usepackage[font=footnotesize]{subfig}
\usepackage{hyperref}
\usepackage{epstopdf}
\usepackage{cleveref}
\usepackage{morefloats}
\usepackage{adjustbox}
\usepackage{hhline}
\usepackage{authblk}
\usepackage{algpseudocode}
\usepackage{algorithm}
\usepackage{authblk}
\usepackage{balance}

\usepackage{url}

\graphicspath{{./Figures/}}

\begin{document}

\title{ODINI : Escaping Sensitive Data from Faraday-Caged, Air-Gapped Computers via Magnetic Fields}


\author{Mordechai Guri, Boris Zadov, Andrey Daidakulov, Yuval Elovici  \authorcr \and Ben-Gurion University of the Negev \authorcr Cyber Security Research Center\authorcr{gurim@post.bgu.ac.il; zadov@post.bgu.ac.il; daidakul@post.bgu.ac.il; elovici@bgu.ac.il}}
\affil{\textbf{Video:} https://cyber.bgu.ac.il/advanced-cyber/airgap\small \authorcr{(gurim@post.bgu.ac.il) }}%

\maketitle

 \section*{Abstract}
 	Air-gapped computers are computers which are kept isolated from the Internet, because they store and process sensitive information. When highly sensitive data is involved, an air-gapped computer might also be kept secluded in a Faraday cage. The Faraday cage prevents the leakage of electromagnetic signals emanating from various computer parts, which may be picked up by an eavesdropping adversary remotely. The air-gap separation, coupled with the Faraday shield, provides a high level of isolation, preventing the potential leakage of sensitive data from the system.
 	In this paper, we show how attackers can bypass Faraday cages and air-gaps in order to leak data from highly secure computers. Our method is based on an exploitation of the magnetic field generated by the computer's CPU. Unlike electromagnetic radiation (EMR), low frequency magnetic radiation propagates though the air, penetrating metal shielding such as Faraday cages (e.g., compass still works inside Faraday cages). Since the CPU is an essential part of any computer, our covert channel is relevant to virtually any device with a CPU: desktop PCs, servers, laptops, embedded systems and Internet of Things (IoT) devices. We introduce a malware code-named 'ODINI' that can control the low frequency magnetic fields emitted from the infected computer by regulating the load of the CPU cores. Arbitrary data can be modulated and transmitted on top of the magnetic emission and received by a magnetic receiver ('bug') placed nearby. We provide technical background and examine the characteristics of the magnetic fields. We implement a malware prototype and discuss the design considerations along with the implementation details. We also show that the malicious code does not require special privileges (e.g., root) and can successfully operate from within isolated virtual machines (VMs) as well. We present signal generation, transmission algorithms, and discuss data encoding and modulation schemes. We analyze the covert channel and evaluate it on various types of computers. Finally, we propose different types of defensive countermeasures such as signal detection and signal jamming to cope with this type of threat.

	\section{Introduction} 
One of the main goals of advanced persistent threat (APT) attacks is to steal sensitive information from compromised organizations. Currently, defending computer networks from APTs and sophisticated cyber-attacks is a complicated task, which involves maintaining multiple layers of security systems. This includes updating protection software in the host computers, configuring firewalls and routers, managing access controls, using centralized credential systems, and so on. Nevertheless, despite a high degree of protection, as long as the local area network in connected to the Internet, a motivated adversary will find a way to breach the network, evade security mechanisms, access sensitive data, and send it out to the attacker side (e.g., Vault 7 \cite{macaskill2017wikileaks} , Sony \cite{zetter2014sony}, and Yahoo \cite{thielman2016yahoo}). 	
		\subsection{Air-Gap Isolation}
When highly sensitive data is involved, an organization may resort to so-called 'air-gap' isolation. In this approach, any type of physical or logical connection between the local network and the Internet is banned. The air-gap separation is maintained by enforcing strict regulations such as forbidding connectivity to unauthorized equipment and hardening the workstations in the network. Today, air-gapped networks are used in military and defense systems, critical infrastructure, the finance sector, and other industries \cite{guri2017bridging,byres2013air}. Two examples of air-gapped networks are the NSANet and the Joint Worldwide Intelligence Communications System (JWICS), classified networks belonging to the United States' Defense Intelligence Agency \cite{Classifi75:online}.
However, even air-gapped networks are not immune to breaches. In the past decade, it has been shown that attackers can successfully penetrate air-gapped networks by using complex attack vectors, such as supply chain attacks, malicious insiders, and social engineering \cite{maybury2005analysis,TrumpPut87:online,abraham2010overview}. For example, in 2017 WikiLeaks published a reference to an hacking tool dubbed 'Brutal Kangaroo,' used to infiltrate air-gapped computers via USB drives \cite{Wikileak92:online}. This tool was used by the attackers to infect Internet workstations of the employees in the organization, and then wait for an employee to insert the USB drive into an air-gapped computer. Using these tools, attackers can breach the network, bypassing security systems such as AVs, firewalls, intrusion detection and prevention systems (IDS/IPS), and the like.

			\subsection{Air-Gap Covert Channels}	
After deploying a malware in the air-gapped network, the attacker may, at some point, wish to retrieve information such as encryption keys, documents, or passwords – a behavior commonly used in APTs. However, despite the fact that infiltration of air-gapped systems has been shown feasible, the exfiltration of data from air-gapped system remains a challenge. Over the years, various out-of-band communication methods to leak data through air-gaps have been proposed. For example, electromagnetic covert channels have been studied for at least twenty years. In this type of communication, a malware modulates binary information over the electromagnetic waves radiating from computer components: LCD screens, communication cables, computer buses, and hardware peripherals \cite{guri2014airhopper,kuhn1998soft,kuhn2002compromising,vuagnoux2009compromising,guri2015gsmem}.
			
				\subsection{Faraday Shielding }
To cope with this type of leakage, particularly electromagnetic leakage, highly sensitive equipment might be placed within metal enclosures known as Faraday shielding or Faraday cages. A Faraday cage is made of conducting material (e.g., wire mesh or metal plates) that shields the area inside the cage from external electric fields. In the context of protecting sensitive equipment, Faraday shields are used to block electromagnetic waves from 1) being leaked from the shielded area or 2) penetrating into it. The most simple case of Faraday shielding is when it is implemented in the computer cabling (e.g., Ethernet, USB, and HDMI cables) to limit their electromagnetic emissions and interferences \cite{hoeft1988measured}. Faraday shields may be deployed in a size of small enclosures to protect entire systems such as desktop PCs and display screens \cite{EMPEMIsh73:online,EMFShiel58:online}, but they may also be used to protect entire rooms and even buildings \cite{wwwcomte48:online}. Faraday shielding renders most air-gap covert channels ineffective, since it prevents the leakage of electromagnetic signals outside to the attacker.

				\subsection{Our Contribution}
In this paper, we present a new type of covert channel that can be used to exfiltrate information from air-gapped computers through Faraday cages. Our method uses low frequency magnetic fields generated by a computer’s CPU.  These fields penetrate metal shields, and hence can be used to bypass the protective Faraday cages.

The following aspects/points represent the contributions of our paper

\begin{itemize}
\item \textbf{Air-gap covert channel.} 	Air-gap covert channel. The communication channel we introduce is an air-gap covert channel. That is, regardless of its ability to bypass Faraday shielding, it is capable of leaking data from disconnected, air-gapped computers.
\item \textbf{Leaking through Faraday shielding.} We introduce a covert channel that can evade the Faraday isolation. That is, it can work in highly secured systems which are kept within a Faraday cages where other types of covert channels (e.g., electromagnetic) fail. As far as we know, this is the first work that discusses the topic of Faraday cages and their evasion concerning covert channels. 
\item \textbf{Bypassing virtual machine (VM) isolation.} Virtual machines are often used as a security measure to add a layer of network isolation between the VM and the external environment. We show that the covert channel works even when the malicious code is executed in an isolated VM.
\item \textbf{Discussion of a magnetic 'bug.'} We introduce the concept of a maliciously implanted magnetic receiver (‘bug’), similar to microphone bugs and radio frequency (RF) receivers used with traditional covert channels. 
\end{itemize}

Our contribution concerning the characteristics of the malware is as follows:
\begin{itemize}
\item\textbf{Hardware availability.} The covert channel uses the standard CPU cores to generate and control the magnetic emanation. This makes the covert channel available on virtually any computer and device with a CPU.
\item\textbf{Stealth.} The transmitting code is considered highly evasive, since it does not perform special CPU instructions or invoke specified API calls. This makes it difficult for AVs and anomaly detection systems to identify the malicious behavior of the malicious transmitter. 
\item	\textbf{Required privileges.} The malware requires no special privileges (e.g., root or admin), and any user-level process can execute it in the system.
\end{itemize}
The paper is structured as follow: In Section \ref{RELATED WORK} we present related work. Section \ref{ATTACK MODEL} describes the attack model. Scientific background on magnetic fields and Faraday cages is provided in Section IV. Signal generation, data encoding, and the transmission protocols are described in Section \ref{TRANSMISSION}. In Section  \ref{ANALYSIS} we present the analysis and evaluation. Countermeasures are discussed in Section \ref{COUNTERMEASURES}, and we present our conclusions in Section \ref{CONCLUSION}.

\section{RELATED WORK}
\label{RELATED WORK}
Conventional covert channels assume the existence of network connectivity between the attacker and the target network. These types of covert channels have been widely studied and discussed in prior academic works and literature. Using covert channels, attackers may hide data within legitimate network traffic (e.g., HTTPS, FTP, and DNS), conceal it in images (steganography), or encode it in packet timings\cite{giani2006data,murdoch2005embedding,zander2007survey}. In cases where there is no direct connection with the target network, the attacker may resort to so-called air-gap covert channels. These covert channels can be classified into five main categories: electromagnetic, magnetic, acoustic, thermal, and optical. 

	\subsection{Electromagentic}
	The electromagnetic based covert channel has been the most researched topic in this field for at least twenty years. In 1988, Kuhn and Anderson \cite{kuhn1998soft} showed how attackers can exfiltrate data over the air-gap by controlling the electromagnetic waves emanating from display screens. In their method, called 'soft tempest,' a malicious code encodes information over AM signals generated by certain dither pixel patterns. Based on this work, in 2001 Thiele \cite{Tempestf48:online} presented an open-source program dubbed 'Tempest for Eliza,' which uses the computer monitor to transmit AM radio signals; the transmissions can be heard from a nearby simple radio receiver. In 2014, Guri et al introduced AirHopper \cite{guri2017bridging,guri2014airhopper}, malware that can leak data from air-gapped networks to nearby mobile phones using controllable electromagnetic signals in the FM radio band emanating from the video cable. Later on, in 2015, Guri et al presented GSMem \cite{guri2015gsmem}, malware that leaks data from air-gapped computers using frequencies in the cellular band emitting from memory buses. In their method, they use a multichannel memory architecture to amplify the transmission power. The transmission is then received by a rootkit placed on baseband firmware of a compromised mobile phone. Researchers also proposed using USB \cite{guri2016usbee}, the data bus, and GPIO ports \cite{funtenna86:online} to generate covert electromagnetic signals for data exfiltration. 
	
\subsection{Acoustic, Optical, and Thermal}	

\textbf{Acoustic.} Exfiltrating data from air-gapped computers via acoustic signals has also been proposed. Hanspach discussed using ultrasonic sound waves (18-22kHz) to transmit data between air-gapped laptops using their speakers and microphones \cite{hanspach2014covert}, however these methods are not relevant when speakers or microphones are not present. In 2016, Guri et al presented Fansmitter \cite{guri2016fansmitter} and DiskFiltration \cite{guri2017acoustic}, two methods enabling exfiltration of data via sound waves when the computers are not equipped with speakers or audio hardware; the binary data is modulated via noise emitted from computer fans and the hard disk drive actuator arm. 

\textbf{Optical.} Data exfiltration using optical signals is another type of covert channel. In 2002, Loughry and Umphress proposed a malicious code that exfiltrates data by blinking the Caps Lock, Num Lock, and Scroll Lock LEDs on the PC keyboard \cite{loughry2002information}. More recently, Guri et al presented a covert channel that uses the hard drive indicator LED \cite{Guri2017} the router LEDs  \cite{guri2017xled} in order to leak data from air-gapped computers and networks. VisiSploit \cite{guri2016optical} is another optical based covert channel in which data is leaked through fast blinking images or low contrast bitmaps projected on the computer screen. Lopes et al 
\cite{lopes2017platform} presented a covert channel based on signals transmitted from IR LEDs in external USB devices attached to the computer. In 2017, Guri et al presented a method that uses the IR LEDs present in surveillance and security cameras to exfiltrate and infiltrate air-gapped networks remotely \cite{guri2017air}. 

\textbf{Thermal.} In 2015, Guri et al presented a thermal based method called BitWhisper \cite{guri2015bitwhisper}. In this technique, an attacker can established bidirectional communication between two adjacent air-gapped computers using heat emissions. The heat is generated by CPU/GPU cores and received by thermal sensors that exist in the PC motherboard.

\subsection{Magnetic }
Magnetic communication by itself is a known topic of research \cite{sojdehei2001magneto}. For example, the MagneLink Magnetic Communication System (MCS) is a system which provides through-the-earth emergency wireless communication based on magnetic fields \cite{2pdf50:online} Near-field magnetic induction (NFMI) communication is another type of magnetic method that allows short range communication between devices\cite{bansal2004near}. These types of communication methods require dedicated magnetic transmitters and receivers, which are not available in the case of our covert channel.
In the context of covert channels, Myhayun suggested using the hard disk drive’s (HDD) magnetic head to generate magnetic emissions, which can be sensed by a nearby smartphone’s magnetic sensor \cite{matyunin2016covert}. The smartphone needs to be located a few centimeters from the transmitting laptop, and the bitrate varies from 0.067 bit/sec to 2 bit/sec. This type of attack is less relevant on a standard workstation where there is a gap between the location of the internal hard drive and the chassis partition. Our method differs from \cite{matyunin2016covert} in the following respects:

\textbf{Signal generation.} We propose to generate the magnetic fields via the CPU which is available on virtually any computerized device today, including devices without magnetic HDDs. This method enables us to control the frequency of the transmissions independently for each CPU core, and hence to be able to use more complex modulation schemes.  

\textbf{Distance and bitrate.} Our covert channel enables higher bitrates and transmission from greater distances, making it more feasible for the proposed attack scenarios.

\textbf{Air-gaps and Faraday shielding.} Our discussion focuses on a covert channel that is relevant to air-gap and Faraday isolation. To the best of our knowledge, this is the first paper that discusses the topic of Faraday cages protection and evasion in the context of covert channels and cyber-security measures. 
Table \ref{table:1} summarizes the existing air-gap covert channels. Table \ref{table:2} summarizes the differences between our work and existing work in the magnetic field.

\begin{table}
	\centering
	\caption{Different types of air-gap covert channels.}
		\begin{tabular}{|l|l|} \hline
			 {\textbf{Type}}& \textbf{Method} \\ \hline
			 \multirow{2}{4em}{Electro-} & AirHopper \cite{guri2014airhopper,guri2017bridging}\\ {magnetic} &GSMem \\ 
			 &USBee \cite{guri2016usbee}\\&Funthenna \cite{funtenna86:online}\\	\hline
			  \multirow{2}{4em}{Magnetic} &ODINI (this paper)\\ &Myhayun (hard disk drive \cite{matyunin2016covert}) \\ \hline
			 \multirow{2}{4em}{Acoustic} &DiskFiltration (Hard disk noise) \cite{guri2017acoustic}\\ &Myhayun Fansmitter (computer fan noise) \cite{guri2016fansmitter} \\ \hline
			 \multirow{1}{4em}{Thermal} &BitWhisper \cite{guri2015bitwhisper}\\ \hline
			  \multirow{2}{4em}{Optical } &Hard drive LED (LED-it-GO) \cite{Guri2017}\\ & VisiSploit (invisible pixels) \cite{guri2016optical} \\ &Keyboard LEDs \cite{Guri2017} \\&Router LEDs \cite{guri2017xled} \\ & \\
			  \multirow{2}{4em}{Infrared (IR)} &aIR-Jumper (security cameras \& infrared) \cite{guri2017air}\\ &Implanted infrared LEDs  \cite{lopes2017platform}\\ \hline			 
		\end{tabular}
	\label{table:1}
\end{table}

\begin{table*}
	\centering
	\caption{Differences between our work and existing work in the magnetic field}
	\begin{tabular}{|l|l|l|l|l|l|}
		\hline
		\textbf{Work}                                                                                         & \textbf{Signal generation}                                                            & \textbf{Transmitters}                                                                     & \textbf{Receiver}  & \textbf{Max bitrate}                                                              & \textbf{Max distance}                                                          \\ \hline
		\begin{tabular}[c]{@{}l@{}}ODINI\\ (this work)\end{tabular}                                           & CPU operations                                                                        & \begin{tabular}[c]{@{}l@{}}PCs, servers,\\  NUK, IoT and,\\ embedded devices\end{tabular} & Magnetic sensors   & 40 bit/sec                                                                        & 100 to 150 cm                                                                  \\ \hline
		\begin{tabular}[c]{@{}l@{}}Hard disk \\ drive ({\cite{matyunin2016covert}})\end{tabular}                                 & \begin{tabular}[c]{@{}l@{}}Magnetic hard\\  disk drive,\\ I/O operations\end{tabular} & \begin{tabular}[c]{@{}l@{}}Laptops with \\ magnetic hard drive\end{tabular}               & Smartphones        & 2 bit/sec                                                                         & 4 to 12 cm                                                                     \\ \hline
		\begin{tabular}[c]{@{}l@{}}Near/far field\\  magnetic,\\ communication {\cite{bansal2004near}} {\cite{2pdf50:online}}\end{tabular} & Magnetic coils                                                                        & Magnetic transmitters                                                                     & Magnetic receivers & \begin{tabular}[c]{@{}l@{}}Tens to hundreds\\  of bits,\\ per second\end{tabular} & \begin{tabular}[c]{@{}l@{}}Two meters\\  to hundreds \\ of,meters\end{tabular} \\ \hline
	\end{tabular}
	\label{table:2}
\end{table*}

\label{ATTACK MODEL}
The adversarial attack model requires running a malicious code in the targeted computer. In addition, there must also be a magnetic receiver hidden near the targeted system. The attack itself consists of four steps: (1) system infection, (2) receiver Implantation, (3) data gathering, and (4) exfiltration. 
\textbf{System infection}. In the initial phase, the attacker infects the target system or network with malware. As discussed, infecting highly secure and even air-gapped networks has been proven feasible in recent years. Note that several APTs discovered in the last decade are capable of infecting air-gapped networks \cite{AndreyNi62:online}, e.g., Turla \cite{TheEpicT20:online}, RedOctober \cite{zaored}, and Fanny \cite{AFannyEq68:online}. As a part of the targeted attack, the adversary may infiltrate the air-gapped networks using social engineering, supply chain attacks, or insiders. \textbf{Receiver Implantation.} The receiver can be a hardware with a magnetic sensor hidden or implanted in close proximity to the target system. An example of such a hardware implant was presented in Snowden's leaked documents. The component, known as COTTONMOTH \cite{COTTONMO81:online}, is a USB connector implanted RF transceiver that attackers used to connect with the air-gapped networks. \textbf{Data gathering}. Having a foothold in the system, the malware starts retrieving interesting data for the attacker. The data might be textual data, encryption keys, credential tokens, or passwords. The data gathered is exfiltrated from a computer – usually a PC workstation or server – which contains the sensitive data to leak. \textbf{Exfiltration}. At the last phase of the attack, the malware starts the data leak by encoding the data and transmitting it via magnetic emanation generated from the computer. The transmissions may take place at predefined times or in response to some trigger infiltrated by the attacker. The leaked data is received by the nearby magnetic receiver and delivered to the attacker encrypted via standard networking (e.g., Wi-Fi).
Note that although the described attack model is complicated, it is not beyond the capability of motivated and capable attackers. Advanced persistent threats coupled with sophisticated attack vectors such as supply chain attacks and human engineering have been shown to be feasible in the last decade. As a reward for these efforts, the attacker can get his/her hands on very valuable and secured information, which is out of reach of other types of covert channels.

	\subsection{Level of Isolation}
\begin{figure}[ht]
		\begin{center}
		
		\includegraphics[width=0.3\textwidth]{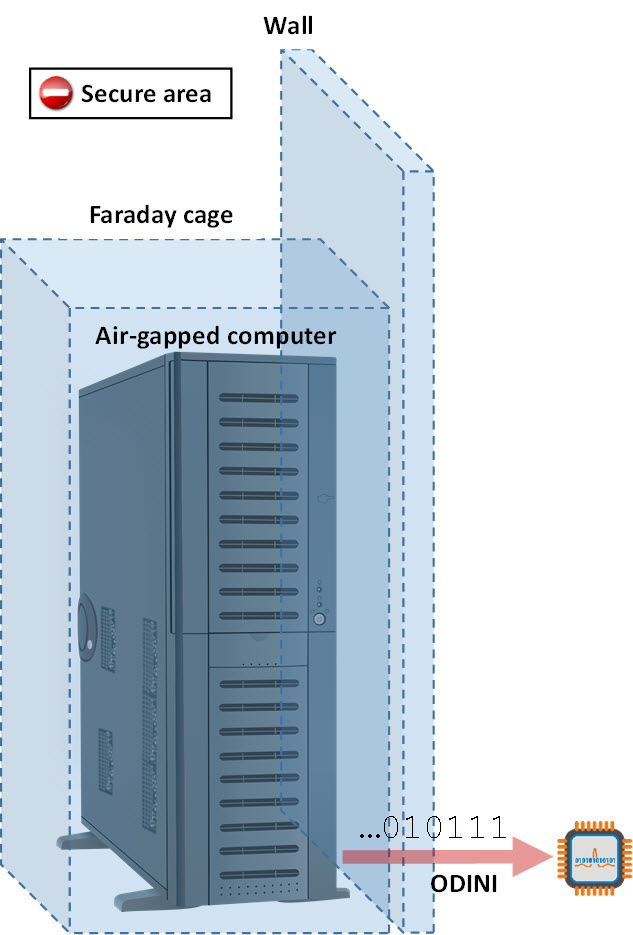}
		\end{center}
	
			\caption{An illustration of the magnetic covert channel (ODINI). Sensitive data is exfiltrated from the secured system, through air-gap and Faraday shielding.}
				\label{fig:figX}
	\end{figure}

	The proposed covert channel is discussed in the context of highly isolated systems which are kept secluded with air-gaps and Faraday cages. However, the attack is also relevant in less restrictive environments as well, for example, computers that are not air-gapped or are not being kept in Faraday cages. In many cases, these computers are highly monitored in the network to detect a potential leakage of data and prevent it. In these cases, the attacker may choose to resort to a type of out-of-band air-gap covert channel, which is not monitored by existing security systems. Figure \ref{fig:figX} illustrates a typical scenario for magnetic exfiltration, in which the magnetic receiver is located near highly secured air-gapped system, which is kept within Faraday shielding.
	
\section{SCIENTIFIC BACKGROUND }	
\label{SCIENTIFIC BACKGROUND}
In this section, we provide the scientific background necessary to understand the magnetic covert channel. We briefly introduce the concept of magnetic fields and discuss Faraday shielding.    
\subsection{Magnetic Field }
Magnetic fields are produced when current flows in a straight wire and are propagated in space at a speed of light. A magnetic field at a given point is specified by its direction and strength and is mathematically represented by a vector field. The international system unit of the intensity for magnetic fields is the tesla (T). One tesla (1T) is defined as the field intensity generating one newton (N) of force per ampere (A) of current per meter of conductor. In practice, a magnetic field of one tesla is very strong and magnetic fields are commonly measured in units of milliteslas ($1mT = 10^{-3}T$) or microteslas ($1\mu T=10^{-9}T$).
Ampère's Law shows that the strength of the magnetic field around an electric current is proportional to the electric current. The strength of the magnetic field is proportional to the third power of the distance from the center of the wire \cite{kodali2001engineering}. The magnetic flux density equation shows that the magnetic field’s rapid decay is proportional to the inverse of the third power of the distance from the source:

\begin{equation}
B(r) = \nabla  \times A = \frac{\mu_0}{4\pi}\left({\frac{3r(m\cdot r)}{|r|}^5}- \frac{m}{|r|^3}\right)
\label{eq:1}
\end{equation}
where $B$ is the strength of the magnetic field in teslas, and r is a distance from the source. The other parameters are the magnetic potential ($A$), magnetic permeability ($\mu_0$), and the magnetic moment ($m$). Note that scientific overview of the magnetic flux density equation is out of the scope of this paper, and we refer the interested reader to textbooks focusing on magnetism \cite{sadiku2014elements}.
As can be seen in \cref{eq:1}, the main disadvantage of the magnetic field is its rapid decay, which limits the distance of magnetic communication compared to that of electromagnetic communication\cite{sadiku2014elements}. In practice, magnetic fields are used for the establishment of short range wireless communication between close devices, a technique commonly referred to as near-field magnetic induction communication \cite{bansal2004near}.

\subsection{Faraday Shielding}
Faraday shielding is an enclosure used to block electromagnetic fields (e.g., radio transmissions) from leaking out or entering into the shielded system. From a scientific point of view, a Faraday shield is a case, which conducts all electromagnetic radiation on its surface. It makes the entire surface to be with equal potential and prevents potential changes inside. Faraday shields may be small in size when protecting computer systems, or very large for protecting entire rooms and laboratories \cite{wwwcomte48:online}. 
Faraday shielding plays an important role in the field of emission security (EMSEC), particularly by providing protection from TEMPEST attacks. In this type of attack, adversaries intercept the electromagnetic radiation emanating from electronic equipment and reconstruct the information processed in the device \cite{kuhn2002compromising}. Faraday shielding copes with this threat by preventing the leakage of electromagnetic signals from the shielded area. Generally, the shielding involves encompassing the device in a Faraday cage that does not permit stray electromagnetic emanations. It should also be noted that there are governmental and commercial standards (e.g., NATO SDIP-27 and NSTISSAM) which require limiting such emanation from devices for security and safety purposes \cite{bindar2014aspects}. 

\subsection{Magnetic Fields and Metal Shielding}
The propagation of electromagnetic and magnetic radiation in conducting mediums such as concrete is better in the low frequencies \cite{kodali2001engineering}, \cite{holloway2000radio}. However, in the case of electromagnetic waves, the antenna required for low frequency transmissions is extremely long, since it is proportional to the wavelength. For example, an efficient transmission of an electromagnetic signal in 100KHz would require an antenna that is more than 3km long.  Magnetic waves, on the other hand, do not depend on antenna length, and hence provide a practical alternative for wireless communication at low frequencies. Specifically, low frequency magnetic waves can propagate through dense medium such as metal, concrete, and soil \cite{kodali2001engineering,2pdf50:online}. 
In the proposed covert channel, we generate magnetic fields at frequencies lower than 50Hz. It is known that low frequency magnetic fields have a low impedance and are difficult to block with metal shielding, since this would require very thick metal surfaces \cite{EMCforSy88:online}. The following analysis shows that magnetic fields at low frequencies can bypass the metal shields of computer chassis and Faraday cages. 
	\begin{figure}[ht]
		\begin{center}
			\includegraphics[width=0.45\textwidth]{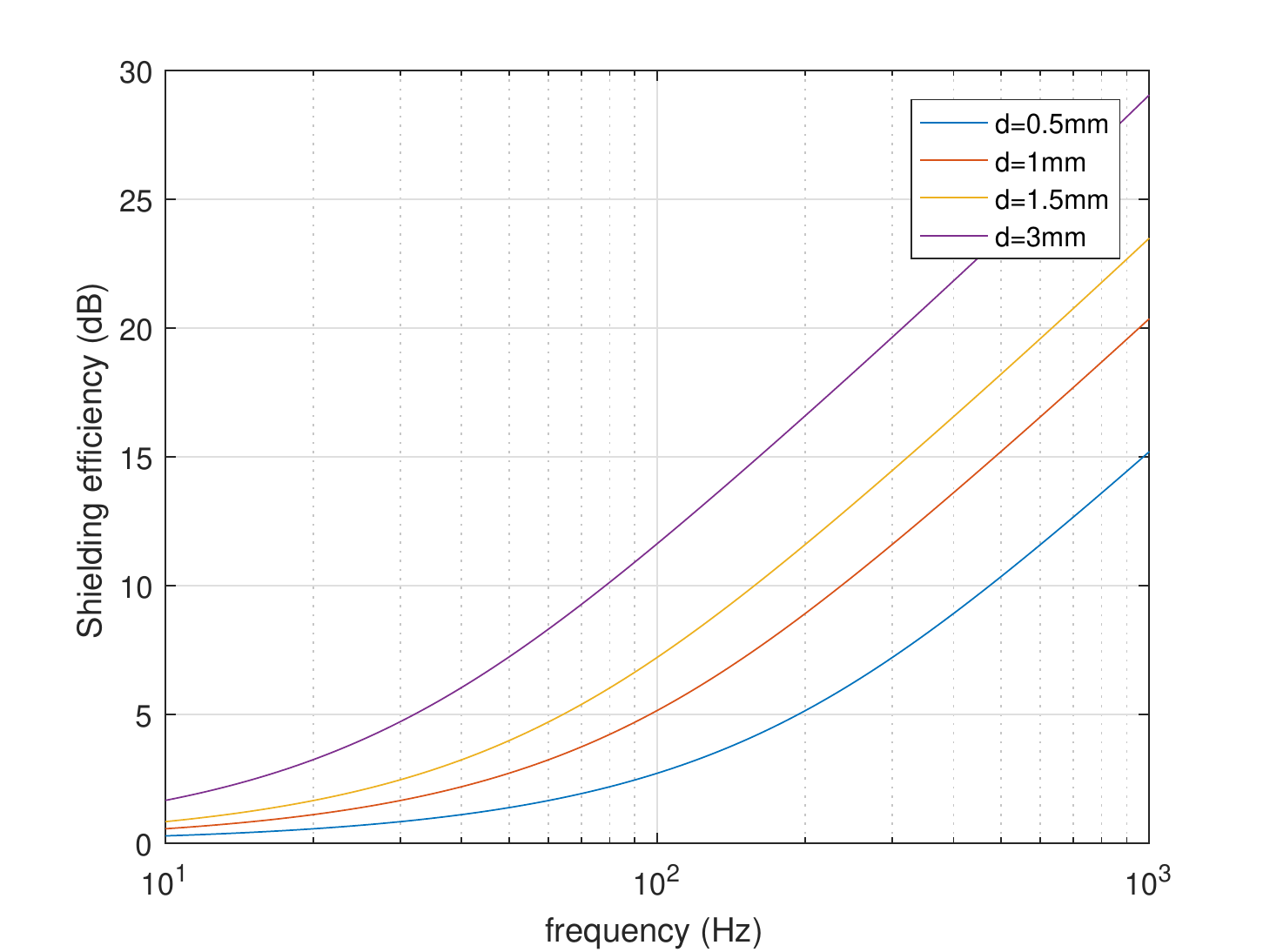}
		\end{center}
	\caption{Shielding efficiency of the closed metal cube shell at a thickness of 0.5mm to 3mm.}
		\label{fig:fig1}
	\end{figure}
Figure \ref{fig:fig1} shows the attenuation of a magnetic field (the reduction in magnitude of magnetic field strength) given a thin walled cube metal shield, based on the shielding approximation formulas \cite{kelha1982design}. The field attenuation is measured in decibels (dB) and is equal to $20 log \frac{E1}{E2}$, where E1 is the field intensity generated on one side of the shield , and E2 is the field intensity received on the other side of the shield. We calculate the efficiency of cubic metal cases at thicknesses of 0.5mm to 3mm in blocking magnetic fields at frequencies below 1000Hz. As can be seen, even thick metal shields are not efficient for low frequencies ($<50$Hz), as the magnetic attenuation is as 5dB at most. The results show that low frequency magnetic fields penetrate the typical computer chassis and Faraday shielding.

\section{TRANSMISSION }
\label{TRANSMISSION}
In this section we describe the signal generation algorithm and present the data modulation schemas and the transmission protocol.

	\subsection{Signal Generation}
	As described in the scientific background section, moving charges in a wire generates a magnetic field. The magnetic field changes according to the acceleration of the charges in the wire. In a standard computer, the magnetic emanation stems primarily from wires that supply electricity from the main power supply to the motherboard. The CPU is one of the greatest consumers of power in the motherboard. Since modern CPUs are energy efficient, the momentary workload of the CPU directly affects the dynamic changes in its power consumption \cite{von2016variations}. By regulating the workload of the CPU, it is possible to govern its power consumption, and hence control the magnetic field generated. In the most elementary case, overloading the CPU with calculations will consume more current and consequentially will generate a stronger magnetic field. Intentionally starting and stopping the CPU workload allows us to generate a magnetic field at the required frequency and modulate binary data over it. 
	We developed a fine-grained approach, in which we control the workload of each of the CPU core independently from the other cores. Regulating the workload of each core separately enables greater control of the magnetic field generated. This approach has the following advantages: 
	
	\begin{enumerate}
	\item Using available cores. Choosing which cores to operate on at a given time, allows us to use only the currently available cores, that is, cores which are not utilized by other processes. This way, the transmission activity won’t interrupt other active processes in the system. This is important for the usability of the computer and the stealth of the covert channel.
	
	\item	Controlling the signal strength. By using different numbers of cores for the transmission, we can control the strength of the magnetic field (e.g., fewer cores consume less power), and hence the amplitude of the carrier wave. This allows us to employ amplitude based modulations in which data is encoded on the amplitude level of the signal. 
	
	\item	Using multiple frequencies. By controlling the workload of each core separately, we can use a different sub-carrier for each transmitting core. This allows us to employ a more efficient modulation scheme such as orthogonal frequency-division multiplexing (OFDM).
	
	\end{enumerate}
	
	To generate a carrier wave at frequency $f_c$ in one or more cores, we control the utilization of the CPU at a frequency correlated to $f_c$. To that end, n worker threads are created where each thread is bound to a specific core. To generate the carrier wave, each worker thread overloads its core at a frequency $f_c$ repeatedly – alternating between applying a continuous workload on its core for a time period of $1/2f_c$ (full power consumption) and putting its core in an idle state for a time period of $1/2f_c$ (no power consumption). 
	
	\begin{figure}[ht]
		\begin{center}
			\includegraphics[width=0.45\textwidth]{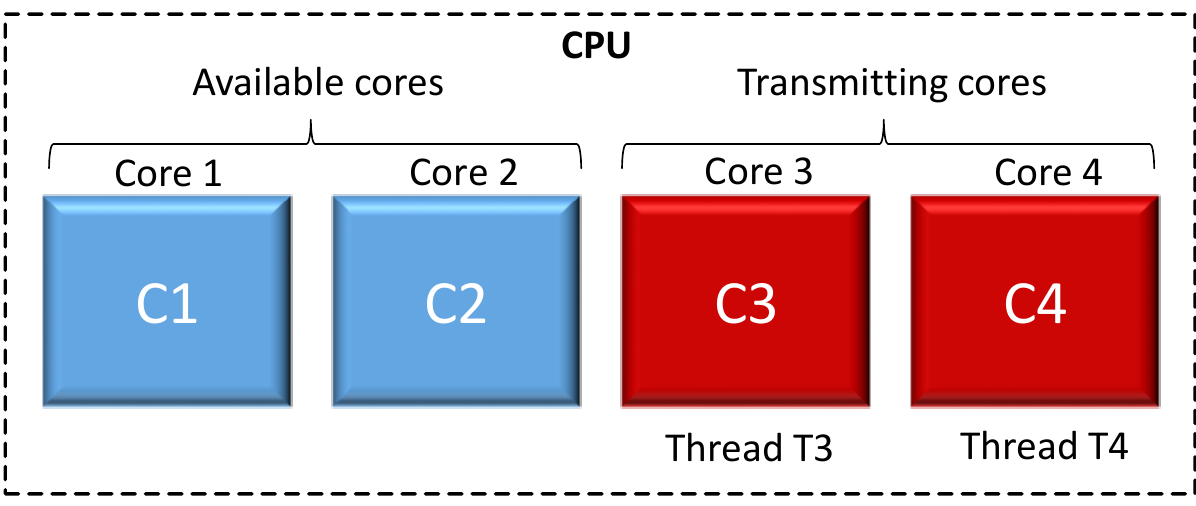}
		\end{center}
	\caption{Signal generation using two threads and two CPU cores.}
		\label{fig:fig2}
	\end{figure}

\begin{algorithm} 
	\caption{WorkerThread (iCore, freq, nCycles0, nCycles1)} 
	\label{alg1} 
	\begin{algorithmic}[1] 
		\State $ bindThreadToCore(iCore) $
		\State $ cycle\_ms \gets 1000/freq $
		\State $ half\_cycle\_ms \gets cycle\_ms*0.5 $

		\While {$ !endTransmission() $}
		\If {thread.data[i] == 0}
		\State $ sleep(nCycles0 * cycle\_ms) $
		\Else
		\For{$j \gets 0\ to\ nCycles1$}
		\State { $ busywait(half\_cycle\_ms) $ }
		\State { $ sleep(half\_cycle\_ms) $ }
		\EndFor
		\EndIf
		
		\State $ i++ $
		
		\EndWhile

	\end{algorithmic}
\end{algorithm}

\begin{algorithm} 
	\caption{busywait(ms)} 
	\label{alg2} 
	\begin{algorithmic}[1] 
		
		\State $ T1 \gets getCurrentTime()$
		\While {$ (getCurrentTime()-T1 < ms)$}
		\texttt{;}
		\EndWhile

	\end{algorithmic}
\end{algorithm}
	

This operation is illustrated in Figure \ref{fig:fig2}, which depicts a system with two worker threads. Threads $T3$ and $T4$ are bound to cores $C3$, $C4$, respectively. Note that cores $C1$ and $C2$ don’t participate in this transmission. When the worker threads $T3$ and $T4$ start, they receive the required carrier frequency $f_c$ and the stream of bits to transmit. The basic operation of a worker thread is described in algorithm 1.

A worker thread receives the core to be bound to (iCore) and the carrier frequency (freq). It also receives the number of cycles for the modulation of logical '0' (nCycles0) and the number of cycles for the modulation of logical '1' (nCycles1). Note that the cycle time is derived from the frequency of the carrier wave (lines 2-3). The thread’s main function iterates on the array of bits to transmit. In the case of logical '0' it sleeps for nCycles0 cycles (line 6). In the case of logical '1' it repeatedly starts and stops the workload of the core at the carrier frequency freq for nCycles1 cycles (lines 8-11). We overload the core using the busy waiting technique as presented in the BusyWait function. This function causes full utilization of the core for the time period and returns. 
Based on the algorithm, we implemented a transmitter for Linux Ubuntu (version 16.04, 64 bit). We used the sched$\_$setaffinity system call to bind each thread to a CPU core. The affinity is the thread level attribute that is configured independently for each worker thread. To synchronize the initiation and termination of the worker threads, we used the thread mutex objects with pthread$\_$mutex$\_$lock() and pthread$\_$mutex$\_$unlock() \cite{pthreadm53:online}. For thread sleeping we used the sleep() system call \cite{sleep3Li5:online}. Note that the precision of sleep() in milliseconds is sufficient given the low frequencies of the carrier waves (e.g., a sleeping cycle of 600ms is required for a 50Hz carrier wave). 
\subsubsection{Stealth}
The transmitting program leaves only a small footprint in the memory, making its presence easier to hide from AVs. At the OS level, the transmitting program requires no special or elevated privileges (e.g., root or admin), and hence can be initiated from an ordinary user space process. The transmitting code mainly consists of basic CPU operations such as busy loops, which do not expose malicious behaviors, making it highly evasive from automated analysis tools.

	\subsection{Data Modulation}
By using different cycle times in the signal generation algorithm, we are able to control the carrier wave frequency. We also have a limited amount of control of the carrier wave’s amplitude by varying the number of cores used for generating the signal. Based on that, we implemented three different data modulation schemes for the transmission: On-off keying (OOK), amplitude-shift keying (ASK), and frequency-shift keying (FSK). We also implement the more efficient orthogonal frequency-division multiplexing (OFDM) modulation scheme. Note that the limited control of the amplitude of the generated carrier wave, also affects the control of the signal’s phase. Thus, we did not use phase-based modulations (e.g., PSK) for this covert channel. In the following sections, we describe each of the four modulations used. We denote the number of cores available for the transmission as $N_c$.	
\subsubsection{On-Off Keying }
In OOK modulation, the data is represented by the presence/absence of the carrier wave.  The presence of a carrier wave represents the symbol '1,' while its absence represents the symbol '0' (Table \ref{table:3}). Note that in our covert channel the amplitude of the carrier wave is unknown to the receiver in advance, and it mainly depends on what type of transmitting computer is used, the number of cores participating in the transmission, and the distance between the transmitter and the receiver. These parameters are synchronized with the receiver during the preamble and bit framing which are described later.

\begin{table}[]
	\centering
	\caption{On-off keying}
	\label{table:3}
	\begin{tabular}{|l|l|l|}
		\hline
		\textbf{Symbol} & \textbf{Carrier wave} & \textbf{\# of cores} \\ \hline
		0               & Present               & $n\leq  N_c$       \\ \hline
		1               & Absent                & $n\leq  N_c$       \\ \hline
	\end{tabular}
\end{table}

\begin{figure}[ht]
	\begin{center}
		\includegraphics[width=0.45\textwidth]{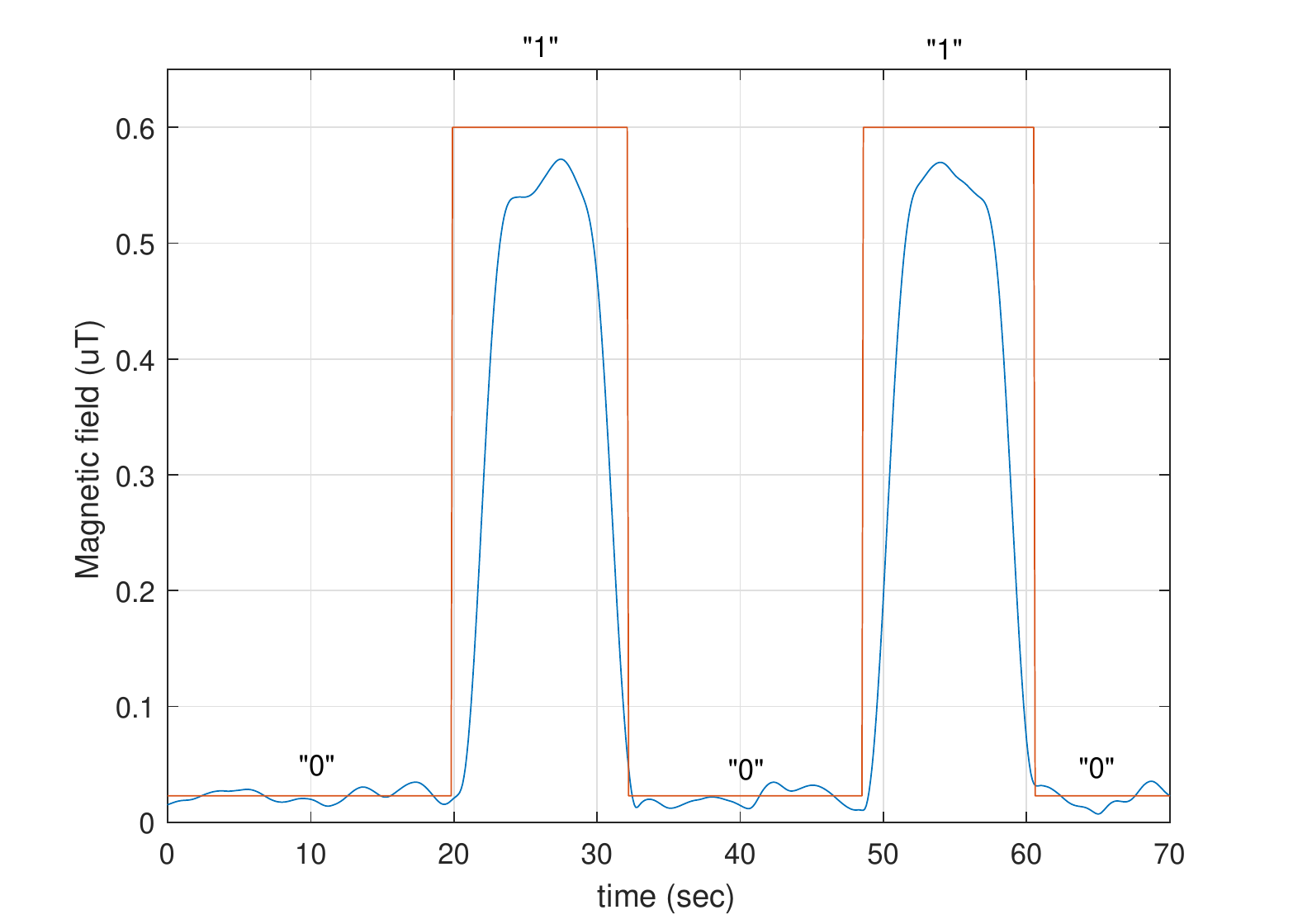}
	\end{center}
	\caption{The waveform of a binary sequence ('01010') modulated with OOK.}
	\label{fig:fig3}
\end{figure}

	Figure \ref{fig:fig3} shows the waveform of a binary sequence ('01010') modulated with OOK and transmitted from a desktop PC with four cores. A magnetic sensor is located 20 cm away from the transmitting computer. The noise level in this case is ~0.02mT, and the carrier wave amplitude is ~0.58mT (SNR of ~29dB).

\subsubsection{Amplitude-Shift Keying  }
In amplitude-shift keying modulation the data is represented by the level of the amplitude of the carrier wave, whereas each level represents a different symbol. The transmitting code controls the signal strength (amplitude) by using different numbers of cores for the transmissions. Accordingly, the number of cores available for the transmission is the number of symbols available. The relationship between the number of symbols available and the number of bits that can be represented by a symbol is 〖M$=2^n$, where $M$ is the number of symbols, and $n$ is the number of bits. Table \ref{table:4} presents a case in which four CPU cores are available for the transmission. We encode the four symbols 00, 01, 10, and 11 by four amplitude levels $A_0$, $A_1$, $A_2$, and $A_3$, respectively.

\begin{table}[]
	\centering
	\caption{Amplitude-shift keying }
	\label{table:4}
	\begin{tabular}{|l|l|l|}
		\hline
		\textbf{Symbol} & \textbf{Amplitude} & \textbf{\# of cores} \\ \hline
		00               & $A_0$               & 0 (or 1)       \\ \hline
		01               & $A_1$                & 2       \\ \hline
		10               & $A_2$                & 3       \\ \hline
		11               & $A_3$                & 4       \\ \hline
	\end{tabular}
\end{table}

\begin{figure}[ht]
	\begin{center}
		\includegraphics[width=0.45\textwidth]{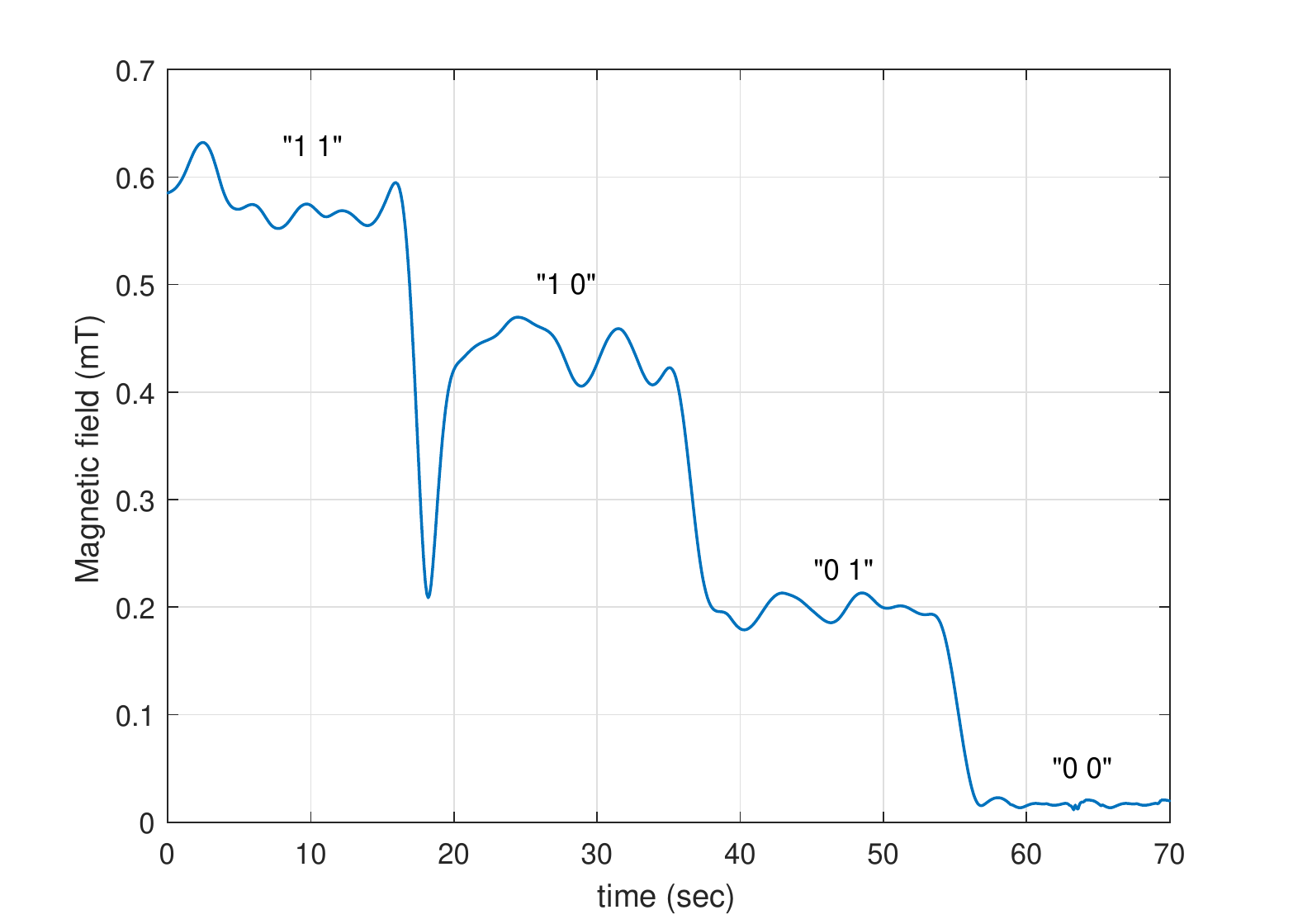}
	\end{center}
\caption{ASK modulation of 111000100 with four amplitudes.}
	\label{fig:fig4}
\end{figure}

Figure \ref{fig:fig4} shows the waveform of a binary sequence ('11100100') modulated with four level ASK and transmitted from a desktop PC with four cores. A magnetic sensor is located at a distance of 20 cm from the transmitting computer. In this case, the symbols '00', '01', '10,' and '11' are represented by four level of amplitudes $A_0$, $A_1$, $A_2$, and $A_3$, and generated by one, two, three, and four cores, respectively.

\subsubsection{Frequency-Shift Keying}
In frequency-shift keying (FSK) the data is represented by a change in the frequency of a carrier wave. Recall that the transmitting code can determine the frequency of the signal by setting the cycle time in the signal generation algorithm. In FSK, each frequency represents a different symbol. Table \ref{table:5} presents a case in which we encode the two symbols '0' and '1' with two frequencies $F_0$, and $F_1$ .

\begin{table}[]
	\centering
	\caption{Frequency-shift keying }
	\label{table:5}
	\begin{tabular}{|l|l|l|}
		\hline
		\textbf{Symbol} & \textbf{Amplitude} & \textbf{\# of cores} \\ \hline
		0               & $F_0$              &$n\leq  N_c$       \\ \hline
		1               & $F_1$              & $n\leq  N_c$       \\ \hline

	\end{tabular}
\end{table}

\begin{figure}[ht]
	\begin{center}
		\includegraphics[width=0.45\textwidth]{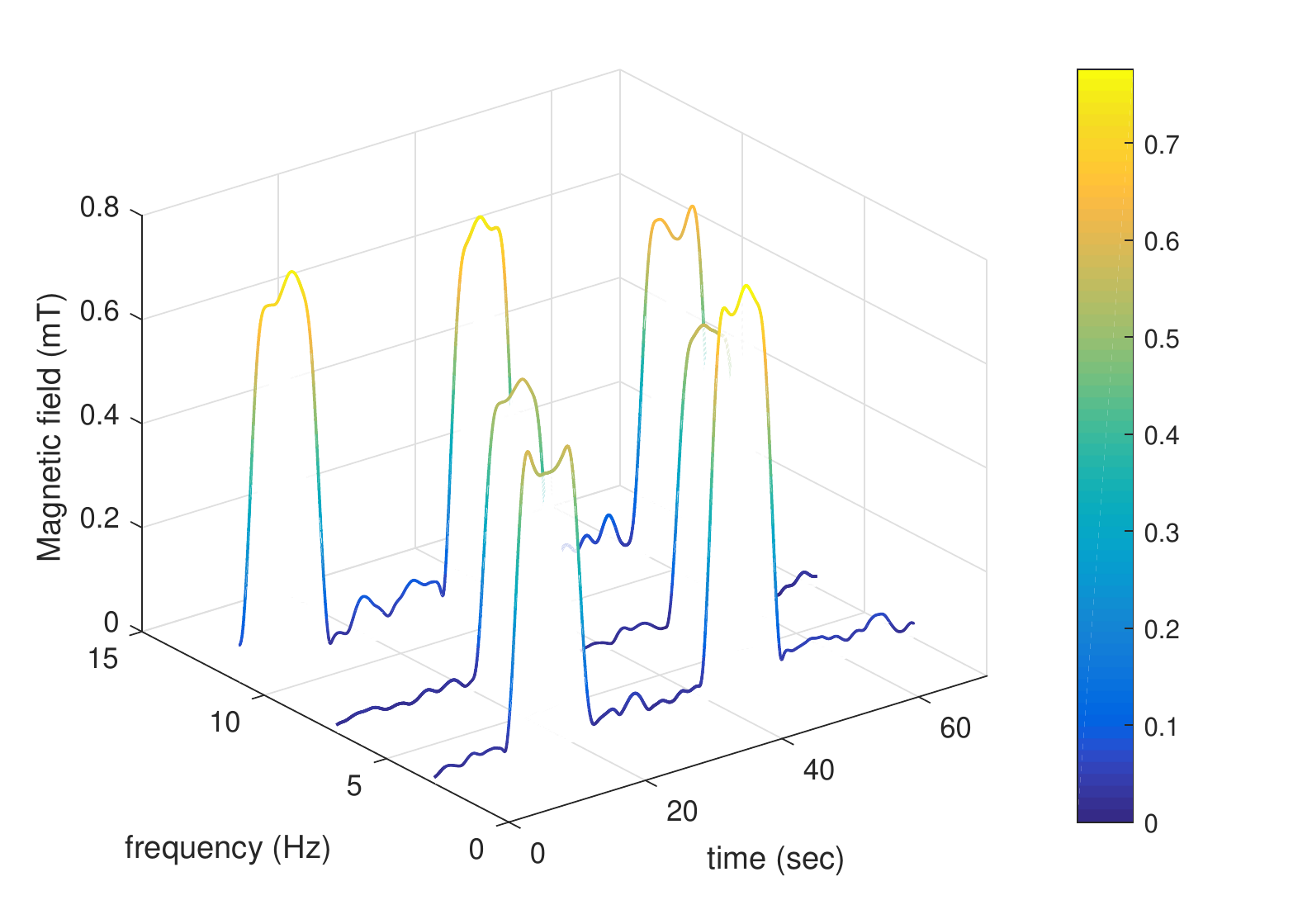}
	\end{center}
\caption{FSK modulation of a binary sequence ('010101010') with three frequencies (3Hz, 7Hz and 13Hz) }
	\label{fig:fig5}
\end{figure}

Figure \ref{fig:fig5} shows the time-frequency spectrogram of a binary sequence ('010101010') modulated with three frequencies FSK as transmitted from a PC with four cores. In this modulation, the frequencies 3Hz, 7Hz and 13Hz have been used to encode 0, 1 and 01 respectively. A magnetic sensor is located at a distance of 20 cm from the transmitting computer.
\subsubsection{Orthogonal Frequency-Division Multiplexing   }
In orthogonal frequency-division multiplexing data is represented by multiple carrier frequencies in parallel. In our case, we use different cores to transmit data in different sub-carriers in a range of 0-50Hz. In each sub-carrier, we used OOK to modulate the data. Note that since the sub-carriers’ signals are generated in parallel, the maximal number of sub-carriers is equal to the number of cores available for the transmissions
$n≤N_c$. For example, in the case of two available cores, we define two sub-carriers (e.g., 7Hz and 13Hz) for the transmission of the data stream. 

\begin{figure}[ht]
	\begin{center}
		\includegraphics[width=0.45\textwidth]{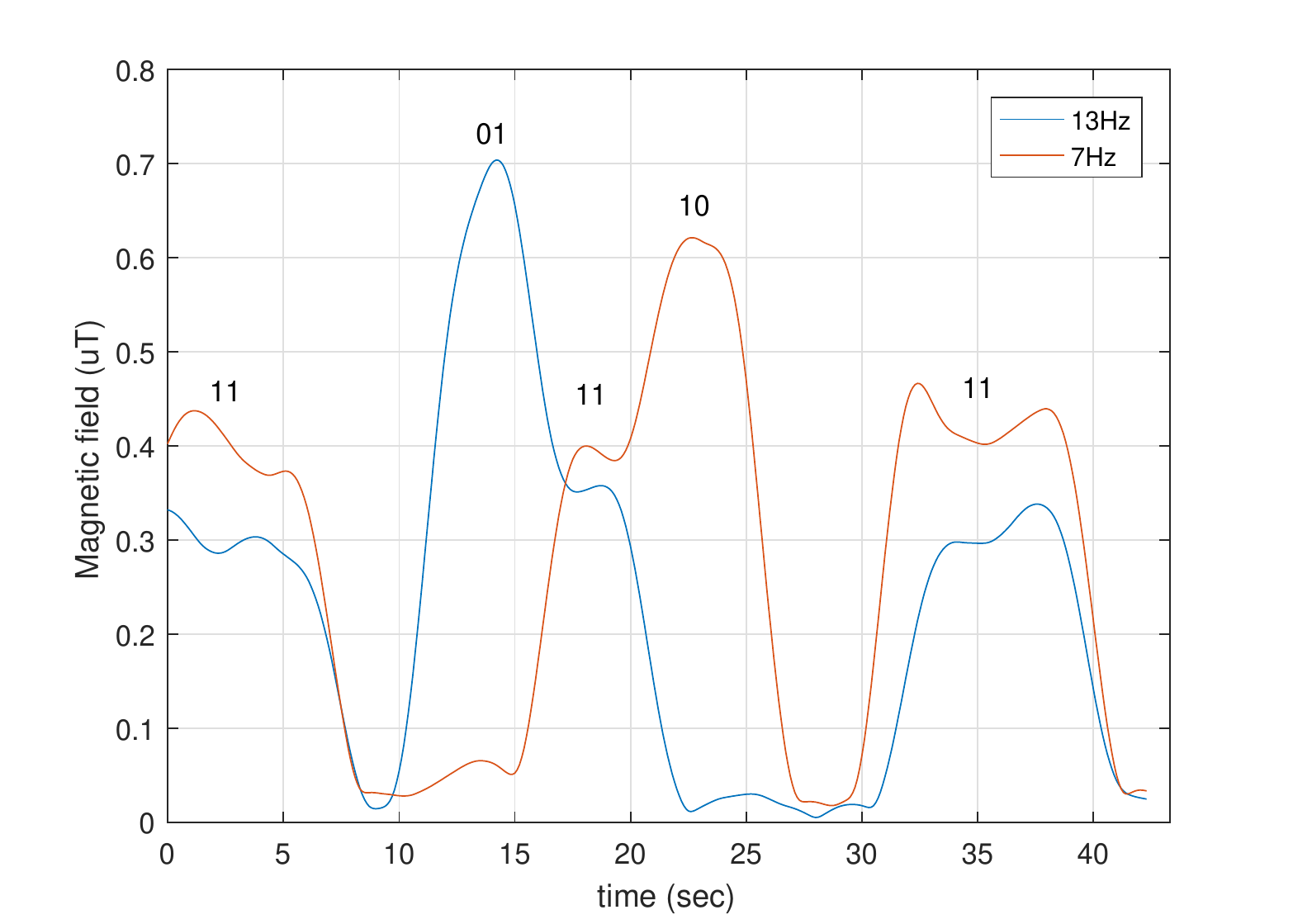} 
			\includegraphics[width=0.45\textwidth]{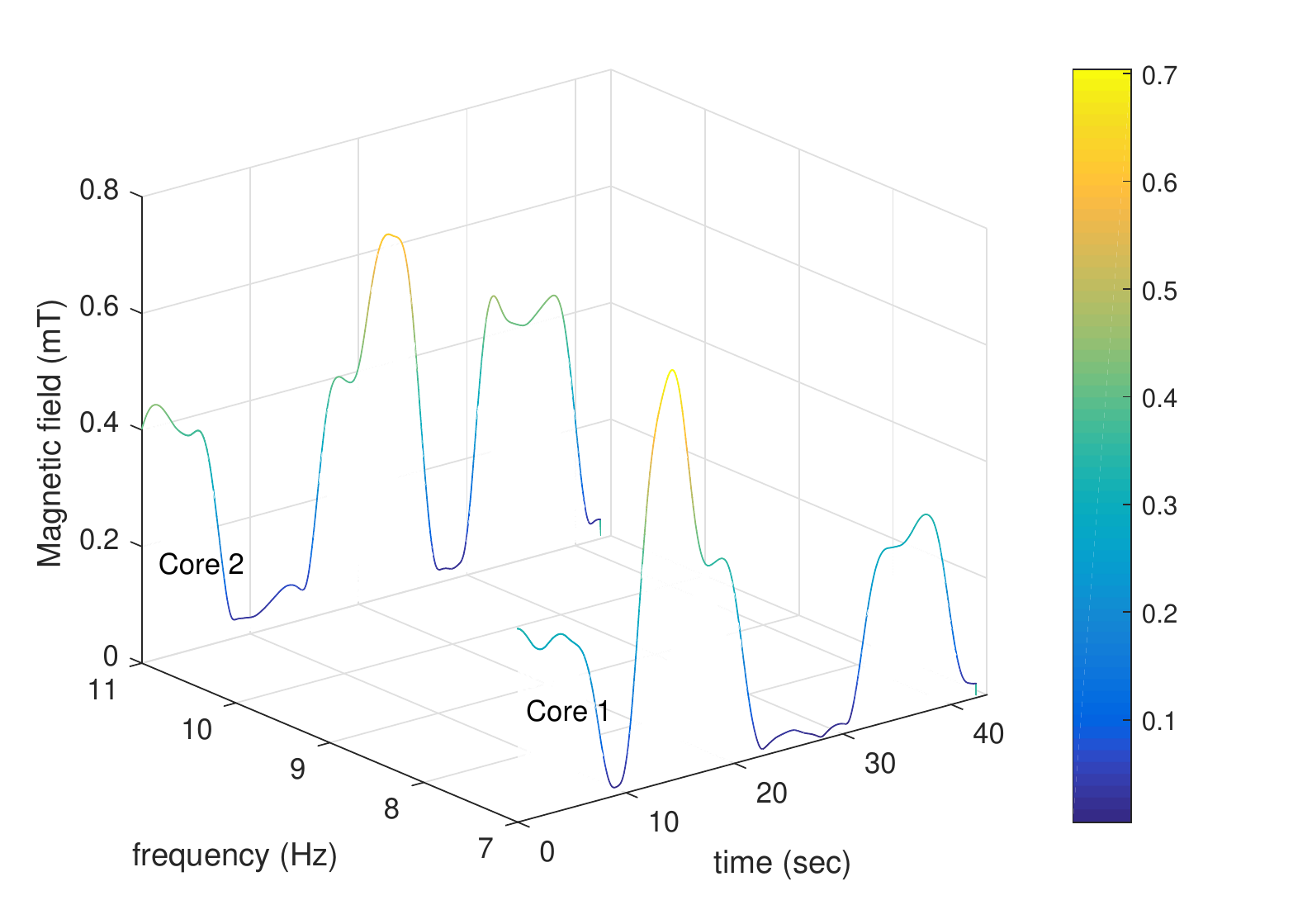} 
	\end{center}
\caption{The waveform and spectrogram of the binary sequence ('1101111011') modulated with OFDM and two sub-carriers (7Hz and 11Hz).}
	\label{fig:fig6}
\end{figure}

Figure \ref{fig:fig6} presents the binary sequence ('1101111011') modulated with OFDM with two sub-carriers as transmitted from a PC with four cores. In this modulation, 7Hz and 13Hz have been used to encode the symbols '00', '01', '10' and '11'. A magnetic sensor is located 20 cm from the transmitting computer.
\subsection{Bit-Framing}
We transmit the data in small packets composed of a preamble, a payload, and a parity bit.  

\begin{itemize}
	\item	Preamble. Like most air-gap covert channels, the unidirectional communication means that the receiver cannot establish a handshake with the transmitter, and hence cannot determine or set the channel parameters before the transmission. To solve this, a preamble header is transmitted at the beginning of every packet. It consists of a sequence of four alternating bits ('1010') which helps the receiver determine the properties of the channel, such as the carrier wave frequency and amplitude. In addition, the preamble allows the receiver to detect the beginning of a transmission and synchronize itself.  
	\item Payload. The payload is the raw data to be transmitted. In our case, we arbitrarily choose 32 bits as the payload size. 
	\item Parity bit. For error detection, a parity bit is added to the end of the frame. The receiver calculates the parity for the received payload, and if it differs from the received parity bit, an error is detected. A more robust protocol may involve advanced error detection and error correction codes (e.g., cyclic redundancy checks). For simplicity we do not consider this in the current paper.
	
\end{itemize}

\section{ANALYSIS \& EVALUATION}
\label{ANALYSIS}
In this section, we present an analysis and evaluation of the proposed covert channel. Note that although near-field and far-field magnetic communication are known research topics[36], [38], such magnetic communication requires specialized and dedicated magnetic transceivers, which are not available in the case of the proposed covert channel. Our experiments focus on the evaluation of the CPU as an unintended, low power magnetic transmitter used for covert communication.

\subsection{Experimental Setup }
\subsubsection{Transmitters (computers)}
The experimental setup consists of six types of computers that are used for the transmissions: two off-the-shelf standard desktop PCs, a laptop computer, a small form factor computer, a server machine with multi-core processors, and a low power embedded device. Unless otherwise specified, the systems in the experiments were run using Linux Ubuntu version 16.04 for 64-bit. A detailed list of the computers is provided in Table \ref{table:6}.

\begin{table*}
	\centering
	\caption{The computers used in the experimental setup }
	\label{table:6}
	\begin{tabular}{|l|l|l|l|l|l|}
		\hline
		\textbf{\#} & \textbf{Name}                    & \textbf{Model}                                                                              & \textbf{Motherboard/board}                                                                                   & \textbf{CPU}                                                                                                        & \textbf{\# of cores}                                                \\ \hline
		\textbf{1}  & PC-1,(Desktop PC)       & \begin{tabular}[c]{@{}l@{}}Infinity \\ desktop PC\end{tabular}                     & \begin{tabular}[c]{@{}l@{}}Gigabyte H87M-\\ D3H\end{tabular}                                        & \begin{tabular}[c]{@{}l@{}}Intel Core i7-4770 CPU\\  @ 3.4GHz\end{tabular}                                 & \begin{tabular}[c]{@{}l@{}}4 (8 \\ threads)\end{tabular}   \\ \hline
		\textbf{2}  & PC-2,(Desktop PC)       & \begin{tabular}[c]{@{}l@{}}Lenovo \\ desktop PC\end{tabular}                       & Panda L-IQ45                                                                                        & \begin{tabular}[c]{@{}l@{}}Intel Core Quad-Q9550 \\ CPU @ 2.83GHz\end{tabular}                             & \begin{tabular}[c]{@{}l@{}}4 (4\\  threads)\end{tabular}   \\ \hline
		\textbf{3}  & Laptop                  & \begin{tabular}[c]{@{}l@{}}HP ProBook \\ 650 G2\end{tabular}                       & Intel                                                                                               & \begin{tabular}[c]{@{}l@{}}Intel Core i5-Q6200U \\ CPU @ 2.4GHz\end{tabular}                               & \begin{tabular}[c]{@{}l@{}}2 (4\\  threads)\end{tabular}   \\ \hline
		\textbf{4}  & Server                  & \begin{tabular}[c]{@{}l@{}}IBM System\\  x3500 M4\end{tabular}                     & Intel C602J                                                                                         & \begin{tabular}[c]{@{}l@{}}Intel Xeon CPU\\  E5-2620\end{tabular}                                          & \begin{tabular}[c]{@{}l@{}}12 (24 \\ threads)\end{tabular} \\ \hline
		\textbf{5}  & NUK,(small form factor) & \begin{tabular}[c]{@{}l@{}}Intel NUK. \\ Lenovo \\ ThinkCentre\\ M93p\end{tabular} & \begin{tabular}[c]{@{}l@{}}Intel Q87 express \\ chipset for \\ ThinkCentre\\  M93/M93p\end{tabular} & \begin{tabular}[c]{@{}l@{}}Intel Core i7\\ -4785T\end{tabular}                                             & \begin{tabular}[c]{@{}l@{}}4 (8\\  threads)\end{tabular}   \\ \hline
		\textbf{6}  & IOT,(IoT/,embedded)     & \begin{tabular}[c]{@{}l@{}}Raspberry\\ Pi 3\end{tabular}                           & \begin{tabular}[c]{@{}l@{}}Raspberry Pi 3 \\ model B V1.2\end{tabular}                              & \begin{tabular}[c]{@{}l@{}}Quad Core Broadcom \\ BCM2837 64-bit ARMv8,\\ processor Cortex A53\end{tabular} & 4                                                          \\ \hline
	\end{tabular}
\end{table*}

Hyper-Threading. Note that modern Intel CPUs support the Hyper-Threading technology \cite{marr2002hyper} . In this technology, each physical core exposes two logical (virtual) cores to the operating system. The CPU shares the workload between the logical cores when possible for better utilization. In the experiments, we bound the transmitting threads to the system's logical cores rather than the physical cores, i.e., in a system with four physical cores and eight virtual cores we can potentially run eight concurrent transmitting threads.

\subsubsection{Recevier   }
For the reception, we used the Honeywell HMR2300 magnetic sensor \cite{Datashee40:online}. This is a digital magnetometer which is capable of sampling the strength and direction of a magnetic field in three axes. It is in use in a wide range of applications such as compassing and navigation, traffic and vehicle detection, laboratory instrumentation, and security systems. The three internal magnetoresistive sensors are oriented in orthogonal directions to measure the X, Y, and Z vector components of a magnetic field. The output is converted to 16-bit digital values using an internal analog-to-digital converter. The sensor resolution of approximately 70nT (nano-teslas) and the sampling rate is up to 154 samples per second. 
\subsubsection{Measurement Setup}
\begin{figure}[ht]
	\begin{center}
		\includegraphics[width=0.45\textwidth]{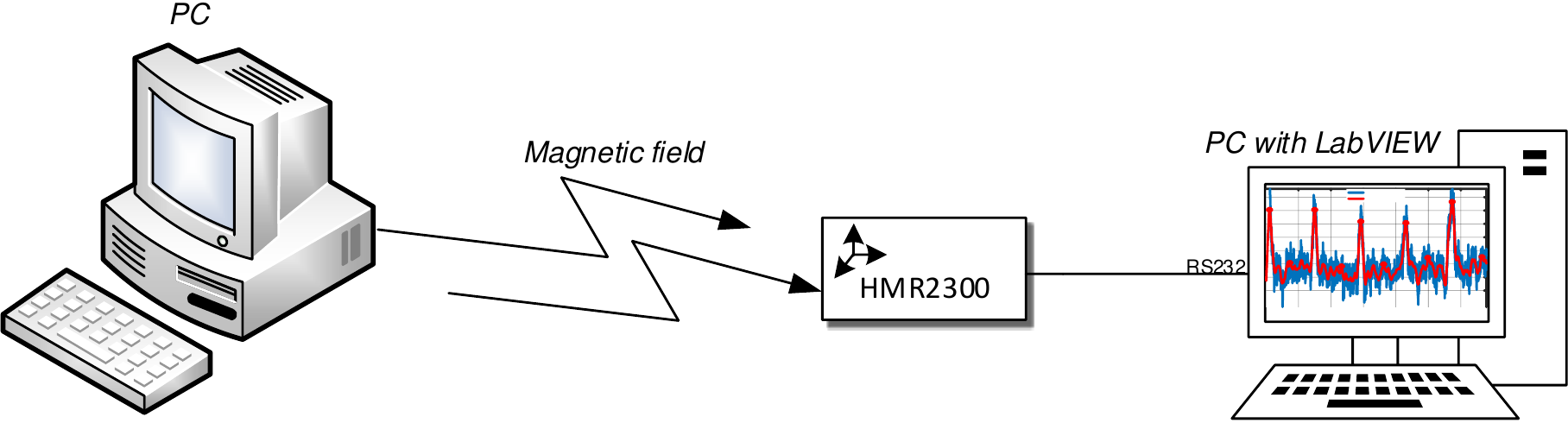} 
	\end{center}
	\caption{The measurement setup}
	\label{fig:fig7}
	
\end{figure}

The measurement setup is shown in Figure \cref{fig:fig7}. The Honeywell HMR2300 is connected to the computer using a serial communication port (RS-232) which is configured to a full-duplex 19,200 data rate. The data is collected with the system driven by a LabVIEW data flow visual programming language. 
\subsection{Signal Strength}

\subsubsection{Number of Cores}
\begin{figure}[ht]
	\begin{center}
		\includegraphics[width=0.45\textwidth]{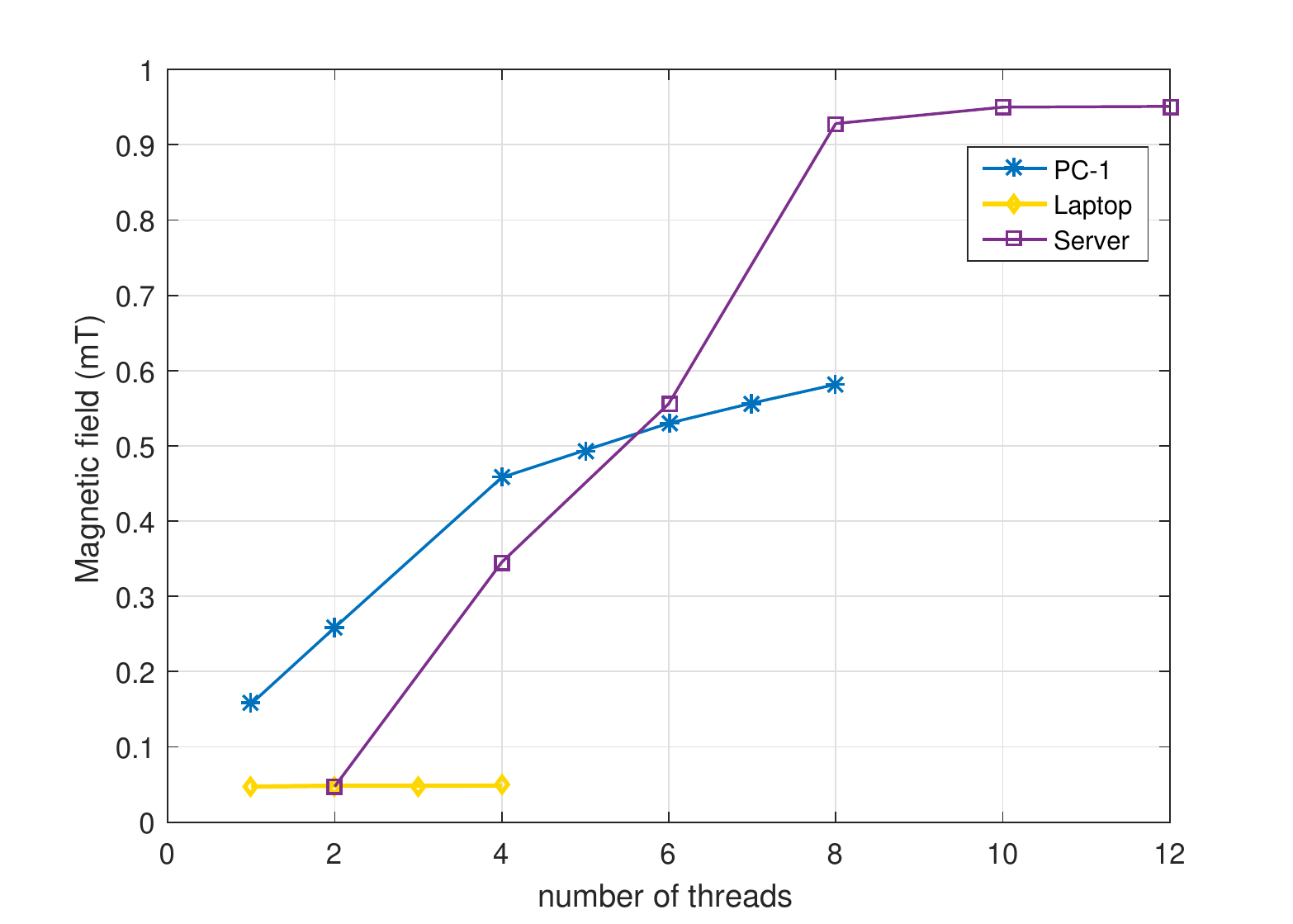} 
	\end{center}
\caption{Measurement of the magnetic signal generated by different number of threads on three computers: PC-1, laptop, and server}
	\label{fig:fig8}
\end{figure}

As discussed, the number of cores used in the transmission directly influences the strength (amplitude) of the magnetic signal (i.e., more transmitting threads yield a stronger signal). Figure \ref{fig:fig8} shows the measurements of three different transmitters: a PC (PC-1), a laptop, and a server. In this test, we used one thread per core and set the carrier frequency to 20Hz. The magnetic field measured at a distance of 20 cm from the transmitting computers showed a gradual increase when an increasing number of threads are used. As expected, the twelve core server showed the greatest increase; from a magnetic field strength of 0.05mT (two threads) to a magnetic field strength of 0.9mT (twelve threads). The magnetic field of the PC increased from 0.15mT (one thread) to almost 0.6mT (eight threads). The magnetic field of the laptop showed almost no increase in the magnetic field strength between one and four threads. 
\subsubsection{Distance   }
\begin{figure}[ht]
	\begin{center}
		\includegraphics[width=0.45\textwidth]{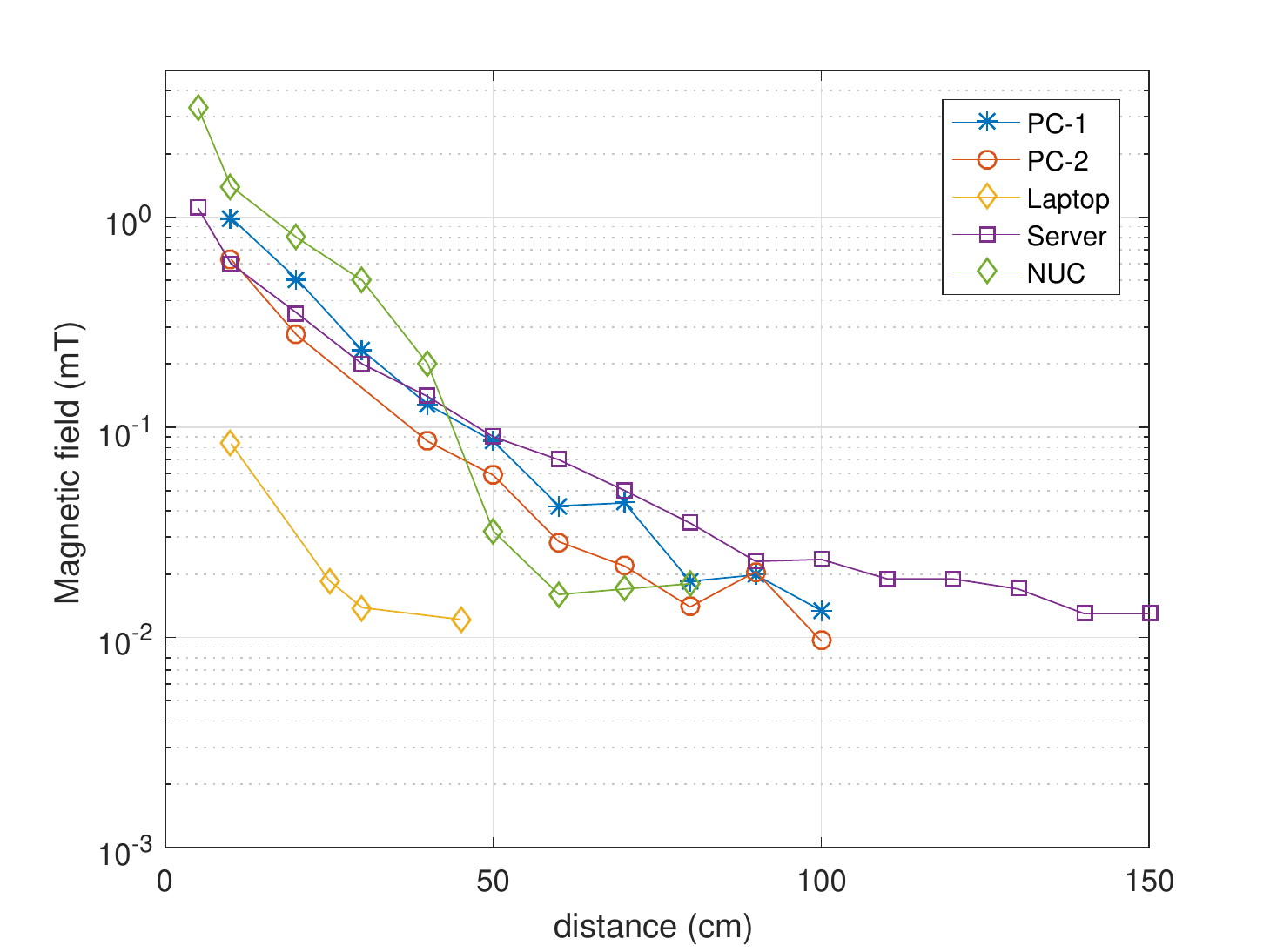} 
	\end{center}
\caption{A 20Hz signal generated by five computers (PC-1, PC-2, laptop, server, and NUK) measured at distances of 0 to 150 cm.}
	\label{fig:fig9}
\end{figure}

The strength of a magnetic field decreases fast, inversely proportional to the third power of the distance from the magnetic source. Figure \ref{fig:fig9} and Table \ref{table:7} show the strength of magnetic signals as measured at various distances from five transmitters. Note that Figure \ref{fig:fig8} shows the magnetic field in a logarithmic scale. Using the HMR3200 sensor, the magnetic signals were received at a maximal distance of 50 cm for the laptop, 100 cm for the desktop PCs and small form factor PC, and 150 cm for the server. Given the resolution of the HMR2300 sensor and the signal to noise ratio (SNR) levels, we stopped the measurements at a field power of $10^{-2}$ mT. Note that the reception of the magnetic signals at greater distances requires more sensitive sensors such as search coil or fluxgate magnetometers \cite{auster2009themis}. 

\begin{table*}
	\centering
	\caption{Measurements of the magnetic field of five transmitters at various distances}
	\label{table:7}
	\begin{tabular}{|l|l|l|l|l|l|l|l|l|l|}
		\hline
		& \textbf{10 cm} & \textbf{20 cm} & \textbf{40 cm} & \textbf{60 cm} & \textbf{80 cm} & \textbf{100cm} & \textbf{120cm} & \textbf{140cm} & \textbf{150cm} \\ \hline
		\textbf{PC-1}   & 0.99 mT        & 0.51 mT        & 0.13 mT        & 0.042 mT       & 0.019 mT       & 0.013 mT       & -              & -              & -              \\ \hline
		\textbf{PC-2}   & 0.63 mT        & 0.27 mT        & 0.059          & 0.022 mT       & 0.01 mT        & 0.009 mT       & -              & -              & -              \\ \hline
		\textbf{Laptop} & 0.084 mT       & 0.019 mT       & 0.012 mT       & -              & -              & -              & -              & -              & -              \\ \hline
		\textbf{Server} & 0.6 mT         & 0.35 mT        & mT 0.14        & 0.07 mT        & 0.035 mT       & 0.026 mT       & 0.019 mT       & 0.017 mT       & 0.013 mT       \\ \hline
		\textbf{NUK}    & 1.4 mT         & 0.8 mT         & 0.2 mT         & 0.016 mT       & 0.013 mT       &                & -              & -              & -              \\ \hline
	\end{tabular}
\end{table*}

\begin{figure}[ht]
	\begin{center}
		\includegraphics[width=0.45\textwidth]{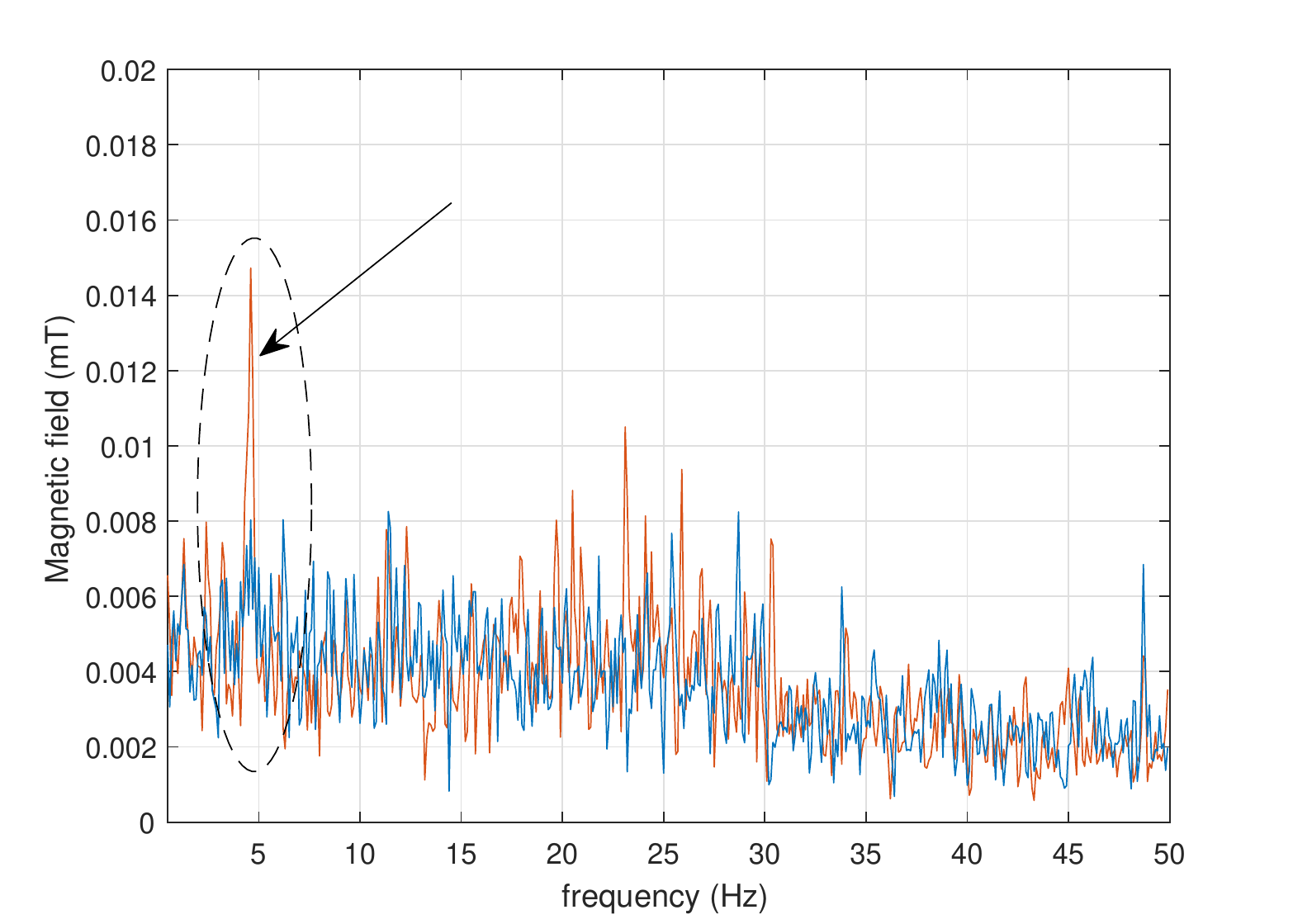} 
	\end{center}
\caption{ A 4.7Hz signal transmitted from Faraday shielded PC-1 as received from a distance of one 100 cm away.}
	\label{fig:fig10}
\end{figure}	

Figure \ref{fig:fig10} depicts the reception of a 4.7Hz signal from PC-1 at a distance of 100 cm. It shows the measured signal in the frequency domain, between 0 and 50Hz. In this case, PC-1 was transmitting at 4.7Hz and located within a Faraday cage, 100 cm from the magnetic sensor. The blue line is shows the background magnetic noise, while the red line shows the sampled signal. A signal strength of 0.014mT is observed at 4.7Hz, which is above the average noise, which is less than 0.008 mT in this band (SNR of ~4.8dB).  		
\subsection{Channel Capacity}
We calculated the maximal bitrate of the covert channel based on the Shannon-Hartley channel capacity limit. Figure 10 shows the channel capacity given the HMR2300 receiver for PC-1, PC-2, laptop, server, and NUK transmitters. In our case, the bandwidth (B) is 50Hz given the sampling rate of the HMR2300 sensor and the low frequencies we used for the transmissions. The S and N were calculated based on the SNR measurements taken for each of the computers at distances of 0 to 150 cm. As shown in Figure \ref{fig:fig10}, for desktop and server computers, the channel capacity varies from ~300 bit/sec to ~30 bit/sec depending on the distance from the computer. The laptop as a transmitter yields a lower channel capacity due to the lower magnetic signals it generates. 
	\begin{figure}[ht]
	\begin{center}
		\includegraphics[width=0.45\textwidth]{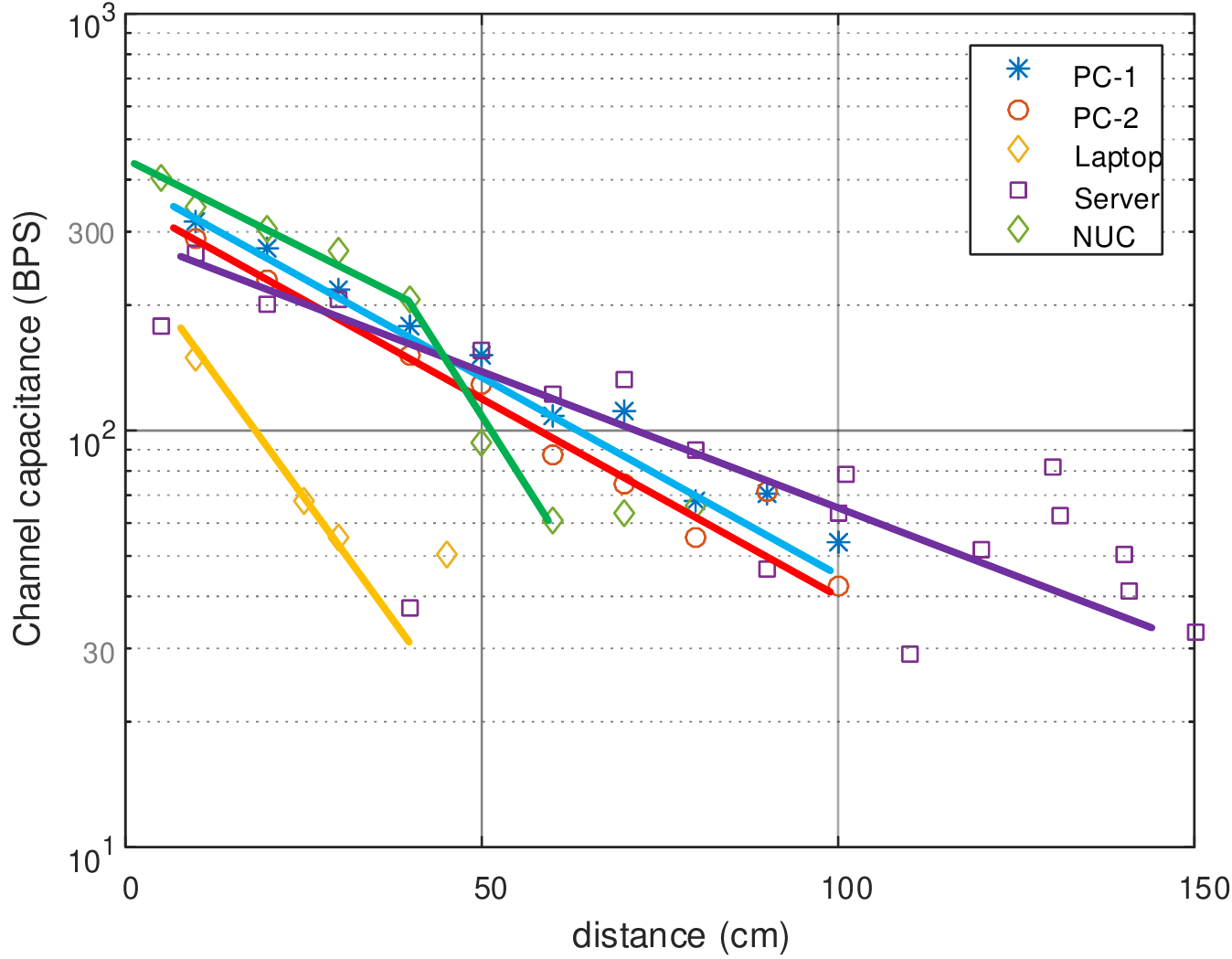} 
	\end{center}
	\caption{The channel capacity of five transmitters based on the SNR measured at a range of distances.}
	\label{fig:fig11}
\end{figure}
	
\subsection{Data Transfer}
The channel capacity represents the upper theoretical limits of a communication channel. The actual bitrate is usually lower than the channel capacity and is determined by the modulation scheme and the quality of the transmitter and receiver used. We measured the bit error rate (BER) of PC-1, the server, and NUK computers for distances of 0 to 120 cm from the transmitting computer. In this test, all of the available cores were used for the data transmission. We tested the transmissions at three bitrates (1, 10, and 40 bit/sec) using the simple OOK modulation and stopped the tests when the results showed a BER of 30$\%$ or higher.  
\begin{table*}
	\centering
	\caption{BER measurements for PC-1, server and NUK}
	\label{table:8}
	\begin{tabular}{|l|l|l|l|l|l|l|l|}
		\hline
		\textbf{PC-1}   & \textbf{5 cm} & \textbf{20 cm} & \textbf{40 cm} & \textbf{60 cm} & \textbf{80 cm} & \textbf{100 cm} & \textbf{120 cm} \\ \hline
		1 bit/sec       & 0\%           & 0\%            & 0\%            & 0\%            & 5\%            & 10\%            & -               \\ \hline
		10 bit/sec      & 0\%           & 0\%            & 25\%           & -              & -              & -               & -               \\ \hline
		40 bit/sec      & 0\%           & 20\%           & -              & -              & -              & -               & -               \\ \hline
		\textbf{Server} & \textbf{0 cm} & \textbf{20 cm} & \textbf{40 cm} & \textbf{60 cm} & \textbf{80 cm} & \textbf{100 cm} & \textbf{120 cm} \\ \hline
		1 bit/sec       & 0\%           & 0\%            & 0\%            & 0\%            & 0\%            & 0\%             & 20\%            \\ \hline
		10 bit/sec      & 0\%           & 0\%            & 28\%           & -              & -              & -               & -               \\ \hline
		40 bit/sec      & 0\%           & 30\%           & -              & -              & -              & -               & -               \\ \hline
		\textbf{NUC}    & \textbf{0 cm} & \textbf{20 cm} & \textbf{40 cm} & \textbf{60 cm} & \textbf{80 cm} & \textbf{100 cm} & \textbf{120 cm} \\ \hline
		1 bit/sec       & 0\%           & 0\%            & 0\%            & 0\%            & 10\%           & 20\%            & -               \\ \hline
		10 bit/sec      & 0\%           & 0\%            & 28\%           & -              & -              & -               & -               \\ \hline
		40 bit/sec      & 0\%           & 30\%           & -              & -              & -              & -               & -               \\ \hline
	\end{tabular}
\end{table*}

The results, presented in Table \ref{table:8}, show that up to a distance of 100 cm, the effective transmission rate is 1 bit/sec for the three computers, with a maximal BER of $10\%$. The higher transmission rates of 10 bit/sec and 40 bit/sec are feasible only when the sensor was in close proximity (5-20 cm away) to the transmitting computer. Note that it is possible to increase the distance by reducing the transmission rates further. However, for the evaluation we consider a transmission rate of 1 bit/sec as the minimal bitrate justifying this attack model. 
	\begin{figure}[ht]
	\begin{center}
		\includegraphics[width=0.45\textwidth]{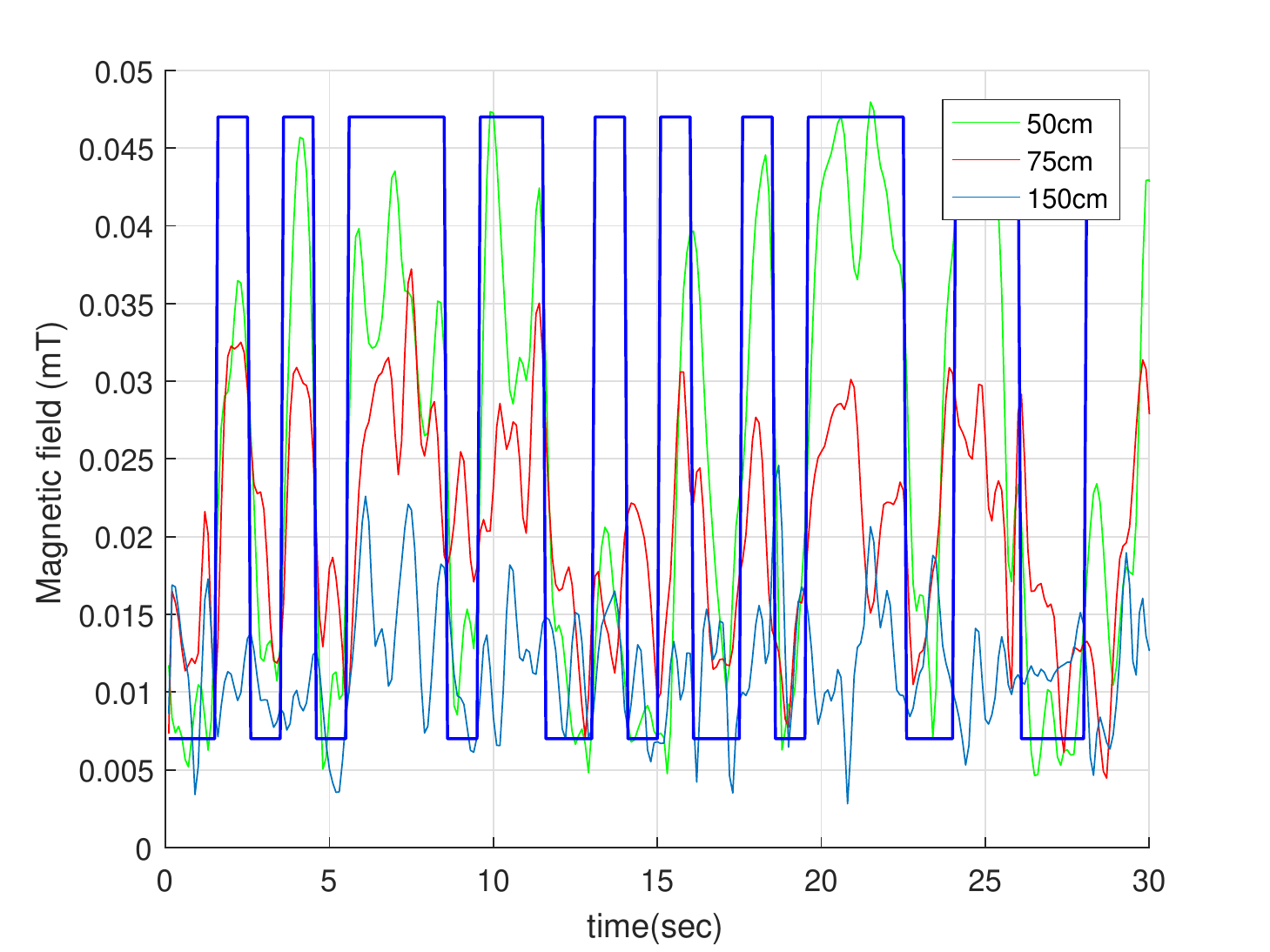} 
	\end{center}
\caption{. Server data measurements for distances of 50, 75, and 150 cm (BER of 0 .}
	\label{fig:fig12}
\end{figure}
Figure \ref{fig:fig12} shows the waveforms of a binary sequence ('10101110110101010111…') encoded in OOK, as transmitted from the server computer. The data was transmitted at a speed of 5 bit/sec and received at distances of 50 cm, 75 cm, and 150 cm away, with an SNR of 10dB, 8dB and 4.4dB respectively.  
It is possible to increase the effective bitrate further by employing advanced signal processing algorithms or by using magnetometers with increased sensitivity and resolution, however we leave the exploration of both of these directions to future work in this field.

\subsubsection{Embedded/IoT Device}
Embedded devices usually consume just a small amount of power, hence emitting lower magnetic fields. Our experiments show that the proposed covert channel also works for such low power devices, when the magnetic sensor is in close proximity of the device. Figure 14 shows the waveform of alternating binary sequence modulated with OOK, as transmitted from the Raspberry Pi 3. The data was transmitted at a speed of 41 bit/sec and received at distances of 10 cm away with a BER of $0\%$ with a SNR of ~15dB.

	\begin{figure}[ht]
	\begin{center}
		\includegraphics[width=0.45\textwidth]{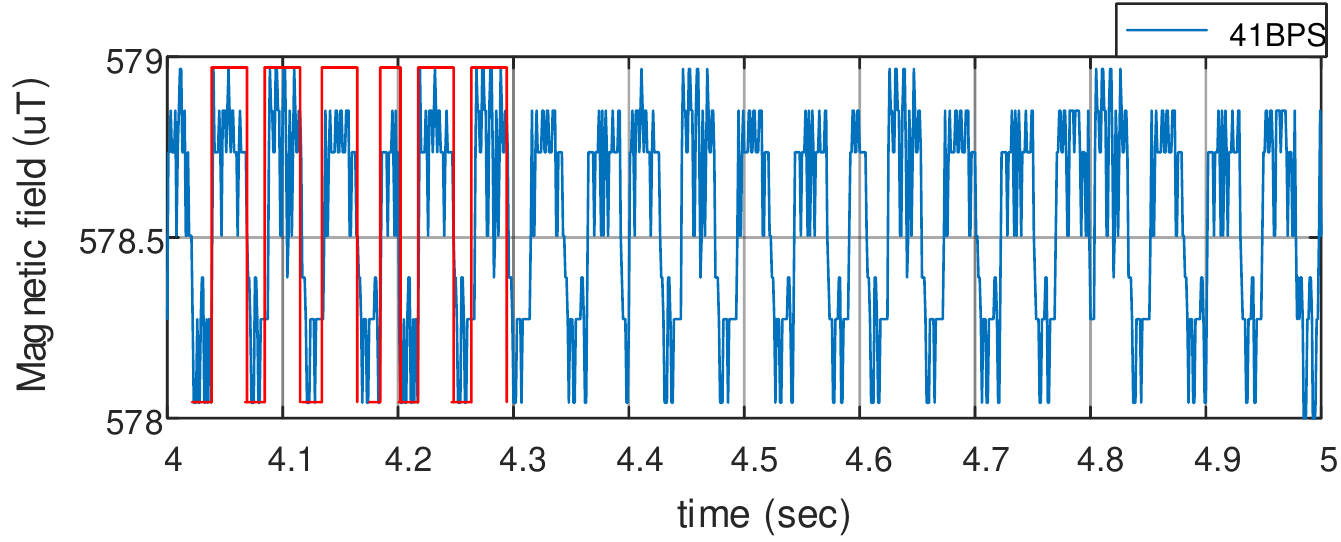} 
	\end{center}
\caption{The waveform of alternating binary sequence modulated with OOK, as transmitted from the Raspberry Pi 3 at 41 bit/sec.}
	\label{fig:fig13}
\end{figure}
\subsection{Virtual Machines}
\begin{figure}[ht]
	\begin{center}
		\includegraphics[width=0.45\textwidth]{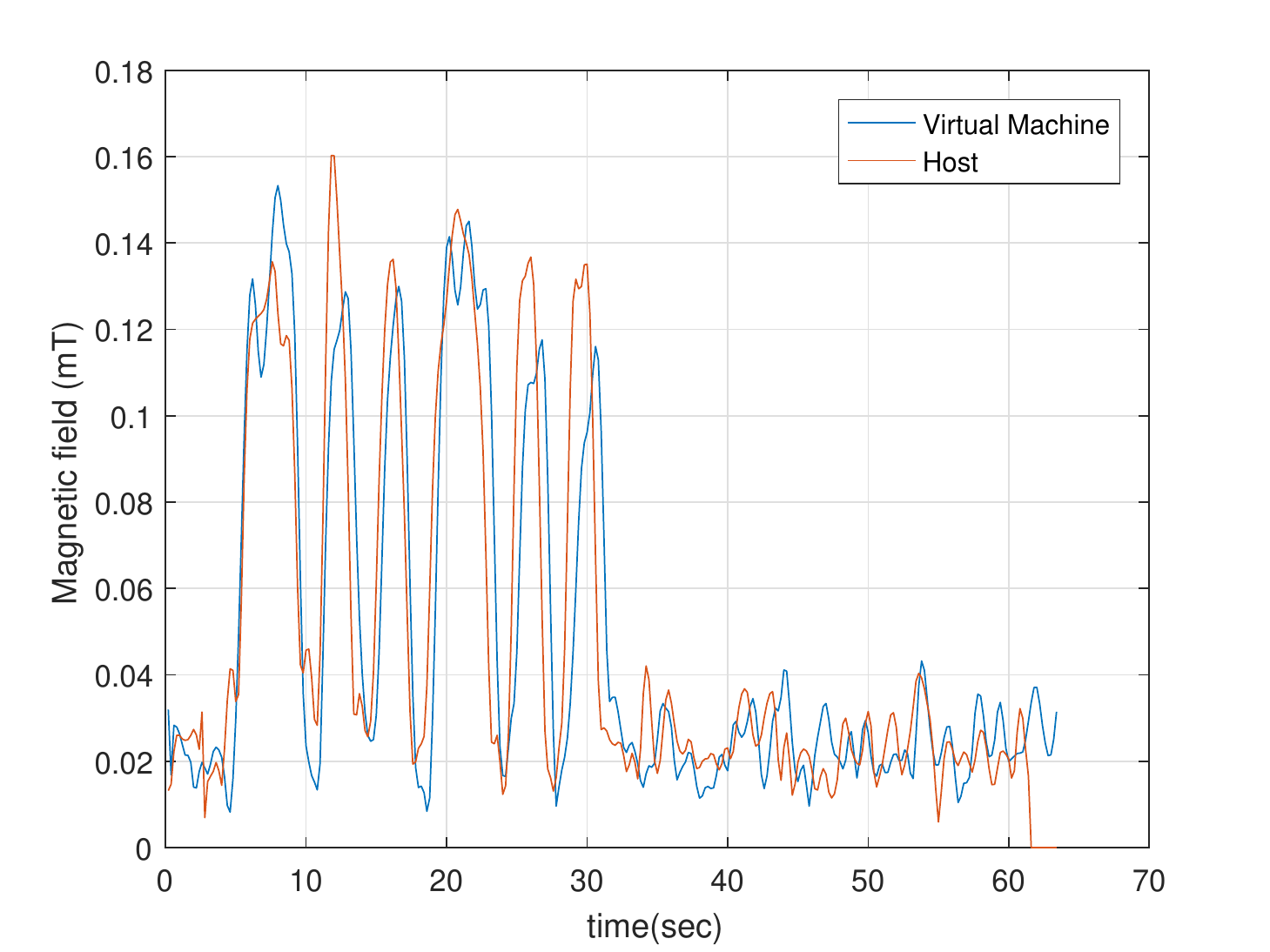} 
	\end{center}
\caption{The waveforms of two signals transmitted from PC-1 (VM/Host).}
	\label{fig:fig14}
\end{figure}

Virtualization technologies are widely used in modern IT environments, including desktop/server virtualization systems and private and public clouds. One of the advantages of virtualization is the resource isolation it provides. Virtual Machine Monitors (VMM) and hypervisors provide a separation between the guest operating system and hardware resources. We examined the operability of a transmitter running in a virtualized environment. Our main goal was to determine whether the virtualization layer caused interruptions or delays which may affect the signal generation.
Figure \ref{fig:fig14} shows the waveforms of two signals transmitted from PC-1. The first signal was generated from the host computer, and the second signal was generated from a VMWare virtual machine. Both signals represent the transmission of the alternating sequence (101010…) using OOK modulation. Both the guest and the host were running Linux Ubuntu 16.04 64-bit. We used VMWare Workstation Player 14.0 for the virtualization and configured the host machine to support four CPU processors. As can be seen, the magnetic signal generated from the VM is highly correlated to the magnetic signal generated directly from the host computer, both having SNR of ~15dB. More specifically, we experienced no time delay or reduction in the power of the signal when it was generated from the VM.

\begin{figure}[ht]
		\begin{center}
			\includegraphics[width=0.45\textwidth]{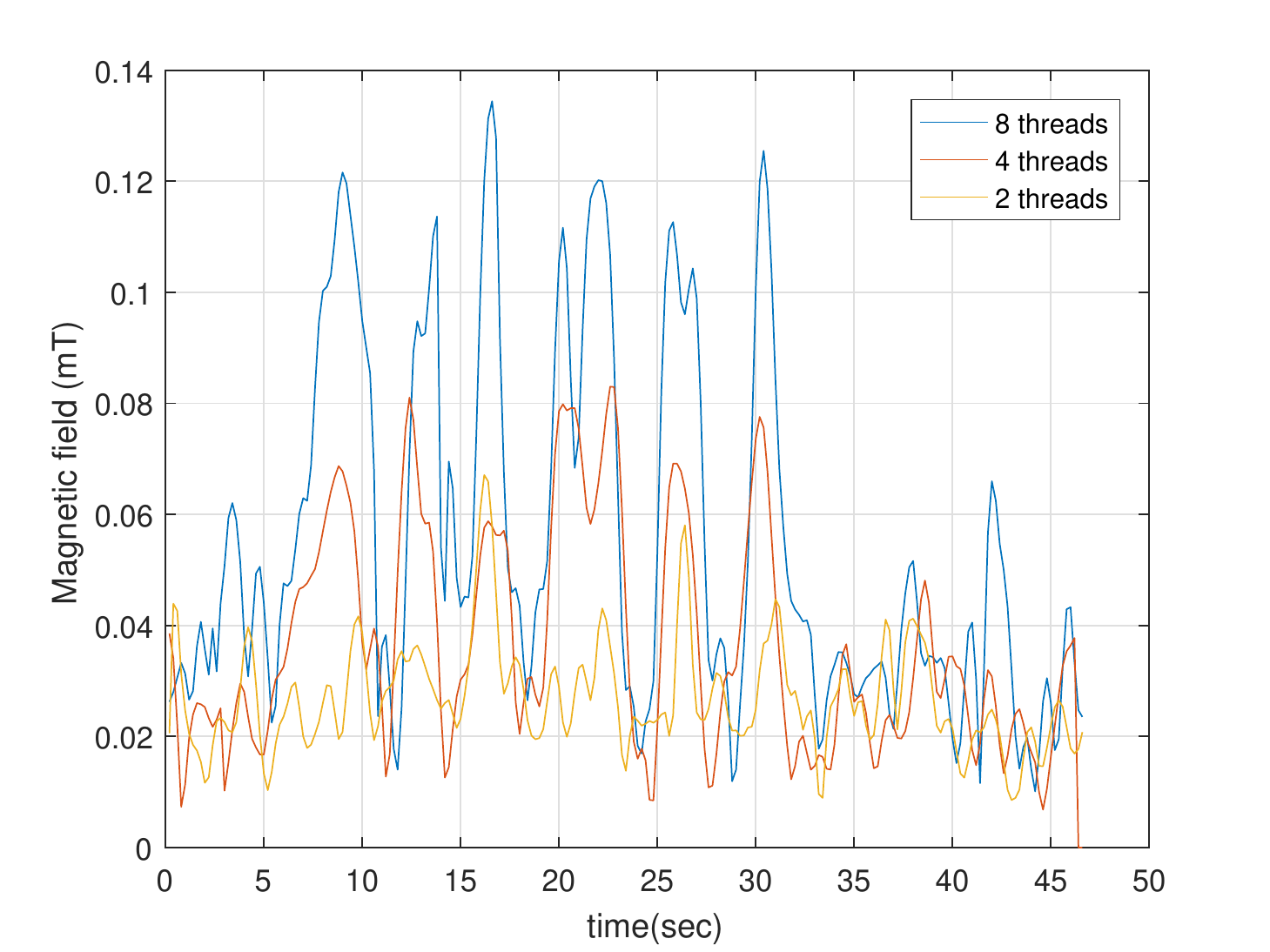} 
	\end{center}
\caption{The waveform of a signal generated from a virtual machine using 2, 4, and 8 threads.}
	\label{fig:fig15}
	\end{figure}

We also investigated the feasibility of controlling the signal strength from virtual machines by using different numbers of cores. Figure \ref{fig:fig15} presents the waveform of a signal generated from a VM with the same setup as described above, using 2, 4, and 8 threads. As can be seen, the transmissions of 2, 4, and 8 threads generate signals at ~0.12mT, ~0.08mT, and 0.04mT, with SNR levels of 15dB, 11dB and 5dB respectively. Generally, employing different numbers of virtual cores in a VM yields different levels of signals, similar to the host-based transmissions. In the context of the communication channel, it allows the attacker to use amplitude modulations from VMs.

\subsection{Interference with Processes }
The threads that generate the magnetic signals share the CPU time with other processes in the operating system. We examined whether the activity of various processes in the system interfere with the signal generation. For this evaluation, we run the transmitting process in PC-1, while employing the following five types of workloads commonly run in desktop PCs:
\begin{enumerate}
\item 	Idle. The system was idle and only the default processes were running in the background.
\item 	Word processing.  The LibreOffice Writer \cite{HomeLibr6:online} was open, and the user typed a document.
\item 	Video playing. The VLC media player \cite{Official45:online} was playing an HD video clip.
\item 	Backup. The Linux rsync \cite{rsync1Li64:online} command was performing a backup of local folders in the HDD.
\item CPU intensive calculations. The Linux matho-primes \cite{UbuntuMa29:online} was performing the calculations of big prime numbers.
\end{enumerate}
Table \ref{table:9} summarizes the SNR measured at a distance of 20 cm from the transmitting computer for each of the five workloads. We used eight threads for the transmission of an alternating bit sequence ('10101010') using OOK modulation. Naturally, the idle state where no other processes interfered with the transmitting process yielded the strongest signal with a measured SNR of 36dB. The word processing and video playing processes consumed just small slices of the CPU time and hence reduced the signal strength at an intermediate level with an SNR of 35 to 36dB. The calculation and backup workloads caused the greatest degradation in the received signals due to the intensive CPU and I/O operations they perform. The SNR in these cases was reduced to levels of 32 to 34dB.

\begin{table*}
	\centering
	\caption{SNR with various workloads}
	\label{table:9}
	\begin{tabular}{|l|l|l|l|}
		\hline
		\textbf{Scenario} & \textbf{Workload} & \textbf{Application/Process} & \textbf{SNRb(dB)} \\ \hline
		\textbf{\#1}      & Idle              & Background processes         & 36.47             \\ \hline
		\textbf{\#2}      & Word processing   & LibreOffice Writer           & 36.00             \\ \hline
		\textbf{\#3}      & Video play        & VLC Player                   & 35.31             \\ \hline
		\textbf{\#4}      & Backup            & rsync                        & 32.04             \\ \hline
		\textbf{\#5}      & Calculations      & matho-primes                 & 33.97             \\ \hline
	\end{tabular}
\end{table*}

The results show that the proposed covert channel is usable even when other active processes are running in the system. However, CPU intensive operations add noise to the generated signal, hence decreasing the effective range and increasing the bit error rate of the transmissions.

\section{COUNTERMEASURES}
\label{COUNTERMEASURES}
\subsection{Detection }
Detection of covert channels could take place by security systems running on the computer. In this approach, security solutions such as AVs, intrusion detection systems (IDSs) and intrusion prevention systems (IPSs) that continuously trace the activities of a computer’s processes and try to detect malicious operations; in the case of a magnetic covert channel, a thread (or group of threads) that abnormally regulates the CPU workload would be tagged as suspicious. However, many types of applications use working threads that affect the processor’s workload, and therefore, such a detection approach would likely suffer from a high rate of false alarms. Another problem in the runtime detection approach is that the signal generation code only involves simple, non-privileged CPU operations (e.g., busy loops), without requiring special instructions or specialized API calls. Tracing non-privileged CPU operations at runtime necessitates entering the processes to a step-by-step mode, which can severely degrade system performance \cite{guri2015gsmem}. Software based detection also suffers from an inherent weakness in that they can be easily bypassed by malware using a wide range of evasion techniques. Another approach is to detect the covert channel externally, by monitoring the magnetic field in the area of the computer. The magnetic field measured is continuously processed to find hidden transmissions or deviations from the standards. Note that finding anomalies in magnetic and electromagnetic spectrums may also suffer from a high rate of false positives \cite{carrara2016air,goher2012covert}.
\subsection{Prevention }

\begin{table*}[!ht]
	\centering
	\caption{List of countermeasures}
	\label{table:10}
	\begin{tabular}{|l|l|l|}
		\hline
		\textbf{Countermeasure}                                                            & \textbf{Description}                                                                                       & \textbf{Cons}                                                                                              \\ \hline
		\begin{tabular}[c]{@{}l@{}}Malicious activity\\  detection (software)\end{tabular} & Detect the transmitting threads                                                                            & \begin{tabular}[c]{@{}l@{}}False positives,\\ Can be bypassed (e.g., by rootkits)\end{tabular}             \\ \hline
		\begin{tabular}[c]{@{}l@{}}Magnetic activity\\  detection (hardware)\end{tabular}  & \begin{tabular}[c]{@{}l@{}}Detect abnormal magnetic \\ field activities\end{tabular}                       & \begin{tabular}[c]{@{}l@{}}False positives,\\ Expensive\end{tabular}                                       \\ \hline
		\begin{tabular}[c]{@{}l@{}}Ferromagnetic\\ shielding (mu-metal)\end{tabular}       & \begin{tabular}[c]{@{}l@{}}Shielding with Ferromagnetic \\ material\end{tabular}                           & Expensive                                                                                                  \\ \hline
		\begin{tabular}[c]{@{}l@{}}Magnetic field \\ jammer (hardware)\end{tabular}        & \begin{tabular}[c]{@{}l@{}}Jamming the transmissions with\\  magnetic field generator\end{tabular}         & Expensive                                                                                                  \\ \hline
		Field cancellation                                                                 & \begin{tabular}[c]{@{}l@{}}Producing the counter magnetic \\ fields\end{tabular}                           & Expensive                                                                                                  \\ \hline
		\begin{tabular}[c]{@{}l@{}}Random workload \\ generator\end{tabular}               & \begin{tabular}[c]{@{}l@{}}Starting random transmissions \\ which jam the transmitter,signals\end{tabular} & \begin{tabular}[c]{@{}l@{}}Degrades system performance,\\ Can be disabled (e.g., by rootkits)\end{tabular} \\ \hline
		Zone separation                                                                    & \begin{tabular}[c]{@{}l@{}}Banning magnetic \\ receivers/electronic equipment\end{tabular}                 & \begin{tabular}[c]{@{}l@{}}Expensive (in terms of \\ physical,space)\end{tabular}                          \\ \hline
	\end{tabular}
\end{table*}

There are three different approaches that can be used to prevent attackers from establishing a magnetic covert channel: shielding, jamming, and zoning.
Shielding. Shielding computers, effectively enclosing them to protect them from low frequency magnetic fields, is considered impractical except for special military or scientific purposes. As discussed in the evaluation section, magnetic fields lower than 50Hz have very low impedance and are difficult to reduce, since this would require very thick metal shielding. For effective magnetic shielding, Ferromagnetic materials such as mu-metal should be used \cite{ter1991improvement}. Ferromagnetic materials require less thick shielding, and hence are more practical for the construction of shielded computer enclosures, however it is difficult to provide effective magnetic shielding against low frequencies even with Ferromagnetic material  \cite{EMCforSy88:online}. Magnetically shielded rooms provide shielding protection on a larger scale. These rooms, which consist of several layers of Ferromagnetic plates, are expensive and weigh several tons. For a more in depth discussion of different approaches for magnetic shielding, we refer the interested reader to \cite{yashchuk_lee_paperno_2013}.
Signal Jamming. Signal jamming is commonly used to mitigate electromagnetic and acoustic covert channels \cite{wilhelm2011wisec}. In this approach, a strong signal that interferes with unauthorized communication is generated in the area requiring protection. The same approach can be used for magnetic communication. Commercial magnetic field generators such as MGA 1030 can generate magnetic fields as strong as 1000 A/m at low frequencies (up to 1kHz) \cite{Schloede39:online}. The power of such a magnetic field is hundreds of times stronger than the magnetic field generated by the CPU, and therefore overrides its magnetic signals. Field cancellation, also known as active magnetic shielding, is another type of signal jamming which is unique to magnetic emanation. This technique uses special equipment that monitors magnetic fields and cancels them by driving a current that produces counter magnetic fields \cite{okazaki2001active}.  
An interesting software level jamming solution is to execute background processes that initiate random magnetic transmissions. The random signals interfere with the transmissions of the malicious process, however random workloads weaken system performance and may be infeasible in some environments (e.g., real-time systems).
Zoning. Procedural countermeasures involve a physical separation of emanating equipment from potential receivers. This approach is referred to as 'zoning' in the National Security Telecommunications and Information Systems Security Advisory Memoranda (NSTISSAM) and NATO standards. For example, the NATO standards SDIP-27 and SDIP-28 define separated zones in which electronic equipment are allowed \cite{anderson2008emission}. In these standards, sensitive computers are kept in restricted areas in which certain equipment is banned. In our case, magnetic receivers of any kind should be banned in the proximity of the sensitive computers.
The detection and prevention based countermeasures and their limitations/weaknesses are summarized in Table \ref{table:10}.

\section{CONCLUSION}
\label{CONCLUSION}
This paper presents a new type of covert channel based on magnetic fields generated by the computer CPU. This method allows attackers to exfiltrate data from isolated, air-gapped computers to a nearby magnetic sensor. Moreover, due to the nature of low frequency magnetic fields, they easily penetrate through metals. This makes our covert channel possible even in a constrained environment where the computers are enclosed within Faraday shielding. We present scientific background and explain the characteristics of magnetic fields and the signal generation process. We introduce a malware codenamed 'ODINI,' which controls the magnetic fields emitted from the computer by controlling the workload of the CPU cores. We show that the malware can work from a user-level process and can operate from within an isolated virtual machine (VM), without requiring special execution privileges. We evaluate the covert channel and show that it works on a wide range of computers and devices. We also propose different types of defensive countermeasures to detect and prevent this threat. Our results show that data can be exfiltrated from air-gap computers via low frequency magnetic fields at bitrates of 1-40 bit/sec. Notably, this type of covert channel can evade Faraday shielding protection, where conventional electromagnetic covert channels fail. 


\bibliographystyle{IEEEtran}
\bibliography{ODINI}
\balance	
\end{document}